\RequirePackage{etoolbox}
\newbool{ipcosubm}
\setbool{ipcosubm}{false}

\ifbool{ipcosubm}{\documentclass[runningheads,envcountsame]{llncs}
}{\documentclass[11pt,letterpaper]{article}
}

\usepackage[T1]{fontenc}
\usepackage[utf8]{inputenc}
\usepackage{lmodern}
\usepackage{amsmath,amssymb,mathtools}
\ifbool{ipcosubm}{}{\usepackage[margin=1in]{geometry}
  \usepackage{amsthm}
}
\usepackage{adjustbox}
\usepackage{microtype}
\ifbool{ipcosubm}{\setlength{\emergencystretch}{1em}}{}
\usepackage[inline]{enumitem}
\usepackage{needspace}
\ifbool{ipcosubm}{\usepackage[backend=biber,style=ext-numeric-comp,sorting=nyt,
    maxcitenames=999,maxbibnames=999,giveninits=true,articlein=false]{biblatex}
  \DeclareFieldFormat{title}{#1}
  \DeclareFieldFormat[article,inproceedings,incollection,inbook]{title}{#1}
  \DeclareFieldFormat{journaltitle}{#1}
  \DeclareFieldFormat{booktitle}{#1}
  \DeclareFieldFormat{pages}{#1}
  \DeclareFieldFormat[article]{volume}{\textbf{#1}}
  \DeclareFieldFormat[article]{number}{(#1)}

}{\usepackage[backend=biber,style=alphabetic,
    maxcitenames=999,maxbibnames=999,maxalphanames=999]{biblatex}
}
\usepackage{xcolor}
\newcommand{\AI}[1]{{\color{green!50!blue}#1}}

\definecolor{linkblue}{RGB}{0,45,110}
\usepackage[colorlinks=true,linkcolor=linkblue,citecolor=linkblue,
  urlcolor=linkblue,pdftitle={Laminar Thin Trees without Cost Loss}]{hyperref}
\usepackage{aliascnt}
\usepackage[capitalise,nameinlink,noabbrev]{cleveref}
\usepackage{xfrac}

\ifbool{ipcosubm}{
  
  \spnewtheorem{claim}[theorem]{Claim}{\bfseries}{\itshape}
  \renewenvironment{proof}[1][\proofname]{\par\normalfont
    \begin{trivlist}
    \item[\hskip\labelsep{\itshape #1.}]\ignorespaces
  }{\qed\end{trivlist}\ignorespacesafterend
  }
}{\newtheorem{theorem}{Theorem}[section]
\newaliascnt{proposition}{theorem}

\aliascntresetthe{proposition}
\newaliascnt{lemma}{theorem}
\newtheorem{lemma}[lemma]{Lemma}
\aliascntresetthe{lemma}
\newaliascnt{corollary}{theorem}
\newtheorem{corollary}[corollary]{Corollary}
\aliascntresetthe{corollary}
\newaliascnt{claim}{theorem}
\newtheorem{claim}[claim]{Claim}
\aliascntresetthe{claim}
\theoremstyle{definition}
\newaliascnt{definition}{theorem}

\aliascntresetthe{definition}
\theoremstyle{remark}
\newaliascnt{remark}{theorem}

\aliascntresetthe{remark}
}

\crefname{theorem}{Theorem}{Theorems}
\Crefname{theorem}{Theorem}{Theorems}
\crefname{proposition}{Proposition}{Propositions}
\Crefname{proposition}{Proposition}{Propositions}
\crefname{lemma}{Lemma}{Lemmas}
\Crefname{lemma}{Lemma}{Lemmas}
\crefname{corollary}{Corollary}{Corollaries}
\Crefname{corollary}{Corollary}{Corollaries}
\crefname{claim}{Claim}{Claims}
\Crefname{claim}{Claim}{Claims}
\crefname{definition}{Definition}{Definitions}
\Crefname{definition}{Definition}{Definitions}
\crefname{remark}{Remark}{Remarks}
\Crefname{remark}{Remark}{Remarks}

\newcommand{\one}{\mathbf{1}}
\newcommand{\cL}{\mathcal{L}}
\newcommand{\PB}[1]{P_{#1}^{\mathrm{base}}}
\newcommand{\PI}[1]{P_{#1}}
\newcommand{\ip}[2]{#1^{\mathsf T}#2}

\DeclareMathOperator{\supp}{supp}

\setlist[enumerate]{label=(\roman*),itemsep=3pt,topsep=5pt}

\usepackage[colorinlistoftodos,textsize=tiny,textwidth=2cm,color=green!50!gray]{todonotes}

\usepackage{xstring}
\usepackage{tikz}
\newcommand{\rz}[2][]{\IfSubStr{#1}{inline}{\todo[color=orange!50,#1]{\footnotesize R: #2}}{\todo[color=orange!50,#1]{R: #2}}}

\title{A Framework for Cost-Preserving Rounding with Applications to Laminar Thin Trees}
\ifbool{ipcosubm}{\titlerunning{A Framework for Cost-Preserving Rounding}
\author{}
\authorrunning{}
\institute{}
\hypersetup{pdfauthor={}}
}{\author{Jannis Blauth\thanks{
Department of Mathematics, ETH Zurich, Zurich, Switzerland.
Email: \href{mailto:jblauth@ethz.ch}{jblauth@ethz.ch}.}
\and
Rico Zenklusen\thanks{
Department of Mathematics, ETH Zurich, Zurich, Switzerland.
Email: \href{mailto:ricoz@ethz.ch}{ricoz@ethz.ch}.}
}
}

\date{}

\begin{document}
\maketitle

\begin{abstract}
In the context of finding thin trees, like laminarly-thin trees or globally-thin trees, it turned out to be challenging to round a fractional solution to an integral one without losing cost, while only having a constant-factor violation of the thinness constraints. 
We provide a general framework for this problem, in a much broader context, that reduces the problem to finding a certain type of rounding algorithm with only an approximate prize guarantee.
We show that, up to a change in constants, the rounding problem with approximate prize guarantee that we use is equivalent to the original question.

As an application, we obtain a polynomial-time algorithm that finds a laminar constantly-thin tree of cost no higher than the one of the natural linear programming solution, thus obtaining the first result without cost loss of this type (even for the special case of chain-constrained trees).
Moreover, our framework gives a new perspective on how to approach the problem of finding globally-thin trees without cost loss.
\ifbool{ipcosubm}{
}{\par\medskip\noindent\textbf{Keywords:} approximation algorithms, thin trees
}
\end{abstract}

\section{Introduction}
\label{sec:introduction}

Finding spanning trees that satisfy additional constraints is a fundamental problem in combinatorial optimization.
A well-studied class of such problems asks to control how often a spanning tree crosses prescribed cuts.
More precisely, given a finite connected undirected graph $G=(V,E)$, a family $\mathcal F\subseteq 2^V$, and bounds $b\in\mathbb R_{\ge0}^{\mathcal F}$, the goal is to find a spanning tree $T\subseteq E$ satisfying
\begin{equation}\label{eq:intro-cut-bounds}
  |T\cap\delta(S)|\le b_S\qquad\forall\; S\in\mathcal F,
\end{equation}
where $\delta(S)\subseteq E$ denotes the edges with exactly one endpoint in $S$.
This setting includes degree-bounded spanning trees, obtained by taking $\mathcal F=\{\{v\}:v\in V\}$, chain-constrained spanning trees, where the members of $\mathcal F$ are nested, and more generally laminar-constrained spanning trees, where $\mathcal{F}$ is laminar, and globally thin trees, where $\mathcal F$ contains all cuts.

A useful approach, also of independent interest, compares the cut loads of a tree to those of a point in the spanning tree polytope $Q\subseteq\mathbb R^E$ of $G$, the convex hull of the characteristic vectors $\chi^T\in\{0,1\}^E$ of its spanning trees.
Given $x\in Q$ and $\kappa\ge1$, a spanning tree $T$ is \emph{$\kappa$-thin for $\mathcal F$ with respect to $x$} if
\begin{equation}\label{eq:intro-thinness}
  |T\cap\delta(S)|\le \kappa\cdot x(\delta(S))\qquad\forall\; S\in\mathcal F.\footnote{The term $x(\delta(S))$ is a shorthand for $\sum_{e\in \delta(S)}x_e$.
    More generally, for $z\in\mathbb R^E$ and $D\subseteq E$, we write $z(D)\coloneqq\sum_{e\in D}z_e$.
  }
\end{equation}
We seek a thinness factor $\kappa$ that is as small as possible, ideally an absolute constant.

Note that \eqref{eq:intro-cut-bounds} is indeed closely related to \eqref{eq:intro-thinness}.
To obtain a tree satisfying \eqref{eq:intro-cut-bounds} up to a factor of $\kappa$, a standard approach is to first use linear programming to find a point $x\in Q$ with $x(\delta(S))\le b_S$ for all $S\in\mathcal F$.
Any tree satisfying \eqref{eq:intro-thinness} then obeys $|T\cap\delta(S)|\le \kappa\cdot x(\delta(S))\le \kappa\cdot b_S$ for all $S\in\mathcal F$, as desired.

There has been significant work and progress on these problems in the last decades.
Not only is it known that there are constant-thin spanning trees for all of the above-mentioned settings (with different constants), but often we can also find such trees in polynomial time with a cost guarantee.
Given costs $c\in\mathbb R^E$ and a fractional point $x\in Q$, we seek a spanning tree $T$ that is thin with respect to $x$ and $\mathcal F$, with cost ideally no larger than $c^\top x$.
We say that an algorithm \emph{preserves cost exactly}, or incurs \emph{no cost loss}, if it guarantees $c(T)\le c^\top x$.
For comparison, a multiplicative cost approximation allows $c(T)\le\rho c^\top x$ for some $\rho\ge1$.
The question we address is how to obtain constant thinness together with exact cost preservation.

Interestingly, so far, among the above-mentioned settings, only the degree-bounded setting has been shown to admit a polynomial-time algorithm that finds a constant-thin spanning tree while also preserving the cost of $x$ exactly. (More precisely, the degree bounds are violated only by at most one unit \emph{additively}.)
For chain constraints, \textcite{OZ18} gave a polynomial-time algorithm achieving thinness nine, without a cost guarantee.
Building on this algorithm, \textcite{LS16} developed a reduction from unweighted to weighted rounding that, for every fixed $\varepsilon>0$, yields a $9(1+1/\varepsilon)$-thin tree of cost at most $(1+\varepsilon)c^\top x$.
For laminar families, \textcite{KO23} gave a polynomial-time algorithm that, for every fixed $\eta>2$ and nonnegative costs $c\in\mathbb R_{\ge0}^E$, returns a $\frac{\eta(2\eta+3)}{\eta-2}$-thin spanning tree of cost at most $\eta c^\top x$.
Thus, the cost factor can be made arbitrarily close to $2$, at the expense of a larger thinness factor.
For globally thin trees, a recent preprint of \textcite{OAI26} gives a deterministic polynomial-time construction, also with a constant factor loss in cost.
This improves earlier superconstant thinness bounds~\cite{AGMOS17,AO15}.
The cost guarantee is obtained by repeatedly finding and deleting thin trees in an auxiliary multigraph, and then selecting the cheapest of the resulting trees.
Thus, these results motivate asking whether one can remove the cost loss while retaining a constant thinness factor.
Notice that letting $\varepsilon$ tend to zero in the chain-constrained guarantee does not achieve this, because its thinness factor grows with $1/\varepsilon$.

In short, there is arguably a lack of techniques for obtaining constant thinness with exact cost preservation.
This is what this paper tries to address, by providing a framework that reduces exact cost preservation to a seemingly weaker rounding task, which we show is in fact equivalent, up to a change in the thinness constant.
Using this framework, we can show how to efficiently obtain laminar constant-thin spanning trees with exact cost preservation, which is the first result of this kind for laminar (and even chain) constraints.
Our framework also provides a natural strategy for getting constant-thin trees without cost loss for other settings, including globally thin trees, even though this question remains open.
Moreover, the framework applies much more broadly than just to thin trees.
We therefore state it in a much broader generality, but then focus on the spanning-tree setting.

\subsection{Our results}

We consider the following setting: given a point $x\in P$ in some polytope $P\subseteq \mathbb{R}_{\ge 0}^n$ and a cost vector $c\in\mathbb R^n$, find an integral point $z\in P$ such that $A z \leq \kappa \cdot b$ for some constant $\kappa \ge1$, while also minimizing $c^\top z$. 
For this setting, we provide a framework to perform this task without cost loss, i.e., we want to fulfill $c^\top z \leq c^\top x$.
The framework reduces exact cost preservation to a seemingly weaker rounding task, which we show is in fact equivalent, up to a change in the constant $\kappa$.
This task asks to round within the smallest face $F_P(x)\subseteq P$ of $P$ containing the input point $x$, while retaining a constant fraction of a given nonnegative prize.
More precisely, given a prize vector $p\in \mathbb{R}_{\geq 0}^n$, besides integrality and constant thinness, we require $z \in F_P(x)$ and
\begin{equation*}
  p^\top z \ge \Omega(p^\top x).
\end{equation*}
Our reduction turns such a rounding algorithm into one that preserves arbitrary linear costs exactly.
Conversely, an algorithm with exact cost preservation yields this face-preserving prize guarantee, even with $p^\top z \ge p^\top x$.
Hence, imposing these additional properties on a rounding procedure does not restrict the class of problems for which exact cost preservation is possible.

In the context of laminar thin trees, we establish the required rounding properties and obtain the following result.
Throughout, we prioritize simple, intuitive techniques and proofs over optimizing constants.

\begin{theorem}\label{thm:intro-laminar}
  For every connected undirected graph $G=(V,E)$, laminar family $\cL\subseteq2^V$, point $x$ in the spanning tree polytope of $G$, and cost vector $c\in\mathbb R^E$, there is a polynomial-time algorithm that computes a spanning tree $T\subseteq E$ satisfying
\begin{equation*}
  c(T)\le c^\top x
  \qquad\text{and}\qquad
  |T\cap\delta(S)|\le 26\cdot x(\delta(S))\quad\forall\; S\in\cL.
\end{equation*}
\end{theorem}

This is the first result achieving constant thinness with exact cost preservation for laminar families, and even for chains.
Because the guarantee holds for any cost function, it also implies that $x$ is a convex combination of constant-thin spanning trees.
Moreover, such a decomposition can be computed in polynomial time; in particular, we can therefore sample a thin spanning tree whose edge marginals are exactly $x$.
(See \cref{app:marginals} for a formal statement and discussion of this standard but important connection.)

Beyond the laminar setting, the framework suggests a strategy for obtaining globally constant-thin trees with no cost loss.
Establishing the required face-preserving prize guarantee in that setting remains open.

\subsection{Further related results}

A generalization by \textcite{STVZ26} of the approach of \textcite{LS16} also provides a strategy for obtaining exact cost preservation from unweighted rounding in certain settings. (\cite{STVZ26} in particular apply their technique in the context of single-source unsplittable flows.)
They obtain exact cost preservation when a face-preserving rounding algorithm has its rounding errors uniformly bounded by a fixed convex set.
This hypothesis does not directly follow from multiplicative cut bounds, whose allowed errors depend on the fractional point being rounded.

\textcite{BKKNP13} gave an iterative relaxation algorithm for spanning trees with laminar cut constraints that preserves the LP cost and violates each cut bound by at most an additive $\mathcal{O}(\log n)$. Our guarantee also preserves the LP cost and gives a constant multiplicative bound even for cuts of small fractional value.

\textcite{NZ24} gave a randomized quasi-polynomial-time algorithm for chain-constrained spanning trees, which extends to laminar families of constant width, that, for every fixed $\varepsilon>0$, violates the cut bounds by at most a factor of $1+\varepsilon$ and returns a tree of cost no larger than an optimum satisfying all bounds exactly.
Their benchmark is the optimum integral solution, rather than an arbitrary given fractional point $x\in Q$.
Apart from not being polynomial-time, and comparing against the integral solution, it heavily relies on near-chain structures to employ a dynamic programming approach and is hard to extend.

\textcite{KOY26} obtained a polynomial-time algorithm that computes a constant-thin spanning tree for all near-minimum cuts of a graph, more precisely, all cuts of size less than $41/40$ times the minimum cut size.
Their result does not provide a cost guarantee.
Contrary to the above-mentioned settings, the cut-family for which they get thinness is controlled by the graph structure rather than being prescribed as input.
\section{Overview of the approach}
\label{sec:approach_overview}

We start by describing a key ingredient underlying our approach.
This is an equivalence between cost-preserving rounding and an auxiliary problem we consider.
It works in the following general context.
Let $P \subseteq \mathbb{R}_{\geq 0}^n$ be a polytope and let $x \in P$ with $Ax \le b$ for some nonnegative constraint matrix $A\in\mathbb R_{\ge0}^{m\times n}$ and right-hand side $b\in\mathbb R_{\ge0}^{m}$.
For a given cost function $c \in \mathbb{R}^n$, we aim to round $x$ to an integral point $y \in P$ with
\begin{enumerate}[nosep]
	\item\label{item:cost_bound} $c^\top y \le c^\top x$, and
	\item\label{item:constraints} $Ay = \mathcal{O}(b)$.
\end{enumerate}
Throughout this work, we assume that $P$ is \emph{solvable}, meaning that we can efficiently optimize linear functions over $P$.
\footnote{Throughout, we write numerical inputs as real numbers. When we say that an algorithm runs efficiently or in polynomial time, we use the standard assumption that these inputs are rational and measure the running time in their binary encoding length. For a polytope $P$ given by an optimization oracle, we also assume a known bound on the encoding length of each inequality in some rational description of $P$. The running time is polynomial in this bound as well.}

Using Lagrangian duality, we show that the following problem with only an approximate prize guarantee implies an algorithm with the above properties:
Find a rounding algorithm returning an integral point $y \in F_P(x)$, where $F_P(x)$ is the minimal face of $P$ containing $x$, such that $Ay = \mathcal{O}(b)$ and $p^\top y \ge \Omega(p^\top x)$ for any given nonnegative prize vector $p$.
A key simplification is the $\Omega$ in the prize guarantee, which only requires a constant fraction of the prize to be preserved.
The constant hidden in the $\Omega$ notation can be chosen arbitrarily, but will impact the constant in the constraint violation.
The statement below formalizes this equivalence, where $\one\in\mathbb{R}^m$ denotes the all-ones vector.

\begin{theorem}\label{lem:general_polytopes}
	Let $n,m \in \mathbb{Z}_{\ge 0}$.
	Let $P\subseteq\mathbb R^n_{\ge 0}$ be a solvable polytope, and let $A\in\mathbb R_{\ge0}^{m\times n}$.
	Let $d \in \mathbb{R}$.
	The following are equivalent.
	\begin{enumerate}
		\item\label{item:weighted} There exists a constant $\gamma>0$ and a polynomial-time
		algorithm that, given $x\in P$ and
		$c\in\mathbb R^n$, returns an integral point $y\in P$
		satisfying
		\begin{equation*}
			c^\top y\le c^\top x, \quad \text{and} \quad Ay\le \gamma Ax +d \one.
		\end{equation*}
		\item\label{item:unweighted} There exist constants $\alpha\in (0,1]$ and $\beta>0$ and a
		polynomial-time algorithm that, given
		$x\in P$ and $p\in\mathbb R_{\ge0}^n$,
		returns an integral point $y\in F_P(x)$ satisfying
		\begin{equation*}
			p^\top y\ge\alpha p^\top x,
			\quad \text{and} \quad Ay\le\beta Ax +d \one.
		\end{equation*}
	\end{enumerate}
	More precisely, given \ref{item:weighted}, then \ref{item:unweighted} holds with $\alpha = 1$ and $\beta =\gamma$. Conversely, given \ref{item:unweighted}, then \ref{item:weighted} holds with $\gamma=\frac{\beta}{\alpha}$.
\end{theorem}

Because it is an equivalence, there is no loss in generality when designing a rounding algorithm for \ref{item:weighted} in \cref{lem:general_polytopes} by instead designing a rounding algorithm for \ref{item:unweighted}.
A proof of \cref{lem:general_polytopes} is given in \cref{sec:general_polytopes}.

For matroid base polytopes, we can simplify and even split these requirements further.
This setting includes the spanning tree polytope, with thinness expressed by $Ax\leq b$, and is how we obtain \cref{thm:intro-laminar}.
Matroid augmentation lets us first find an independent set retaining a constant fraction of the prize and then extend it to a basis.
Moreover, passing to an auxiliary matroid eliminates the explicit face-preservation requirement $y\in F_P(x)$.
We explain these points next.

\subsection{Simplifications for matroids}

We now focus on the case where $P$ is a matroid base polytope.
For a matroid $M=(N,\mathcal{I})$,\footnote{A matroid $M=(N,\mathcal{I})$ is a tuple, where $N$ is the \emph{ground set}, and $\mathcal{I}\subseteq 2^N$ is a nonempty family of subsets of $N$, called \emph{independent sets}, satisfying
	\begin{enumerate*}
		\item \emph{hereditary property:} if $I\in \mathcal{I}$ and $J\subseteq I$, then $J\in \mathcal{I}$, and
		\item \emph{augmentation property:} if $I,J\in \mathcal{I}$ and $|I|<|J|$, then there exists an element $e\in J\setminus I$ such that $I\cup\{e\}\in \mathcal{I}$.
	\end{enumerate*}
	The rank function $r_M\colon 2^N\to\mathbb Z_{\ge0}$ of $M$ is defined as $r_M(S)=\max\{|I|:I\subseteq S, I\in \mathcal{I}\}$ for all $S\subseteq N$.
	We use the standard assumption that a matroid is given by a polynomial-time independence oracle, which is an algorithm that, given $S\subseteq N$, decides whether $S\in \mathcal{I}$.
	We defer the interested reader to \textcite{Schrijver03} for further information about matroids.
}
 let 
\[
\begin{aligned}
\PB{M}
&\coloneqq \operatorname{conv}\{\chi^B : B \text{ is a basis of } M\} \\
&= \left\{
x \in \mathbb{R}_{\ge 0}^N :
\begin{array}{ll}
	x(S) \le r_M(S) & \forall S \subseteq N, \\
	x(N) = r_M(N) &
\end{array}
\right\}
\end{aligned}
\]
denote the corresponding matroid base polytope.
Here, $r_M$ is the rank function of $M$, and $x(S)=\sum_{e\in S}x_e$.

When seeking a rounding algorithm for $\PB{M}$ satisfying \ref{item:unweighted} in \cref{lem:general_polytopes}, we can exploit a useful property of matroid base polytopes to get rid of the requirement that the returned solution is in $F_{\PB{M}}(x)$, the minimal face of $\PB{M}$ containing $x$.
More precisely, any face of a matroid base polytope is itself a matroid base polytope for (possibly) another matroid $M'$.
The matroid $M'$ is a \emph{refinement} of $M$, which means that its base polytope $\PB{M'}$ is obtained from $\PB{M}$ by turning some inequalities of the description of $\PB{M}$ into equalities.

This observation allows us to eliminate the requirement of being face-preserving in \ref{item:unweighted} in \cref{lem:general_polytopes} by moving to the refinement $M'$ of $M$ with $\PB{M'}=F_{\PB{M}}(x)$.
Indeed, because $\PB{M'}=F_{\PB{M}}(x)$, the point $x$ is in the relative interior of $\PB{M'}$ and we therefore only need to return any base of $M'$ to be face-preserving as required by \ref{item:unweighted}.

Moreover, for matroids we can further simplify and split the required rounding guarantees in \ref{item:unweighted} in \cref{lem:general_polytopes} into two independent tasks.
For $p\in\mathbb R^N_{\ge0}$, an algorithm with the guarantees in \cref{lem:general_polytopes}~\ref{item:unweighted} is equivalent to two independent rounding algorithms\AI{:} one that computes an independent set $I$ of $M'$ with
\begin{equation*}
	p(I) = \Omega(p^\top x),
	\quad \text{and} \quad A \chi^I = \mathcal{O}(Ax),
\end{equation*}
and a second algorithm that computes a basis $B$ of $M'$ with
\begin{equation*}
	A \chi^B = \mathcal{O}(Ax).
\end{equation*}
Indeed, we can extend $I$ to a basis $B'$ of $M'$ by adding elements from $B$.
This results in the following equivalence.
\ifbool{ipcosubm}{\Cref{sec:general_matroids} proves this equivalence by combining the two algorithms through matroid augmentation and using refinements to preserve the required face.
}{We provide a formal proof in \cref{sec:general_matroids}.
}

\begin{theorem}\label{lem:matroid_equivalence2}
	Let $m \in \mathbb{Z}_{\ge 0}$.
	Let $M$ be a matroid on a ground set $N$ given by a polynomial time independence oracle.
	Let $A \in \mathbb{R}_{\ge 0}^{[m] \times N}$.
	The following are equivalent.
	\begin{enumerate}
		\item\label{item:weighted_matroid2} There exist constants $\gamma>0$ and $d\in\mathbb R$ and a polynomial-time
		algorithm that, given $x\in \PB{M}$ and
		$c\in\mathbb R^N$, returns a basis $B$ of $M$
		satisfying
		\begin{equation*}
			c(B)\le c^\top x, \quad \text{and} \quad A\chi^B \le \gamma Ax +d\one.
		\end{equation*}
		\item\label{item:unweighted_matroid2}
		There exist constants $\alpha\in (0,1]$, $\beta_1>0$, and $d_1\in\mathbb R$ and a
		polynomial-time algorithm that, given a refinement $M'$ of $M$, $x\in \PB{M'}$, and $p\in\mathbb R_{\ge0}^N$,
		returns an independent set $I$ of $M'$ satisfying
		\begin{equation*}
			p(I) \ge\alpha p^\top x, \quad \text{and} \quad A \chi^I \le\beta_1 Ax +d_1\one,
		\end{equation*}
		and there exist constants $\beta_2>0$ and $d_2\in\mathbb R$ and a
		polynomial-time algorithm that, given a refinement $M'$ of $M$ and
		$x\in \PB{M'}$,
		returns a basis $B$ of $M'$ satisfying
		\begin{equation*}
			A \chi^B \le\beta_2 Ax +d_2\one.
		\end{equation*}
	\end{enumerate}
	More precisely, \ref{item:weighted_matroid2} implies \ref{item:unweighted_matroid2} with $\alpha=1$, $\beta_1=\beta_2=\gamma$, and $d_1=d_2=d$. Conversely, \ref{item:unweighted_matroid2} implies \ref{item:weighted_matroid2} with $\gamma=(\beta_1+\beta_2)/\alpha$ and $d=d_1+d_2$.
\end{theorem}

\cref{lem:matroid_equivalence2} gives a clear strategy for designing rounding algorithms for matroid base polytopes with controlled constraint violation and no cost loss.
We next show how this strategy can be realized to obtain such an algorithm for laminar-thin trees.
As mentioned in the introduction, \Cref{lem:matroid_equivalence2} also provides a strategy to obtain globally thin trees with no cost loss, which remains open.

\subsection{Laminar thin trees}

We now focus on realizing \cref{lem:matroid_equivalence2}~\ref{item:unweighted_matroid2} for the graphic matroid to prove \cref{thm:intro-laminar}.
Hence, let $G=(V,E)$ be an undirected graph, $x$ be a point in its spanning tree polytope, and $\mathcal L\subseteq 2^V$ be a laminar family.
The goal is to find a spanning tree $T\subseteq E$ that is $26$-thin with respect to $x$ on every cut in $\mathcal L$, that is,
\begin{equation*}
    |\delta(L)\cap T|\le 26 \cdot x(\delta(L)) \qquad\forall L\in\mathcal L,
\end{equation*}
without increasing the cost of the solution, i.e., $c(T)\le c^\top x$.
This corresponds to \cref{lem:matroid_equivalence2}~\ref{item:weighted_matroid2} for $M$ being the graphic matroid on $G$, and $A$ specifying the cut constraints for sets in $\mathcal{L}$.

We show the following lemmas, which provide the two independent rounding algorithms required in \cref{lem:matroid_equivalence2}~\ref{item:unweighted_matroid2} for the graphic matroid, thus proving \cref{thm:intro-laminar}.

\begin{lemma}\label{lem:thin_trees_prize_preserving}
	Let $G=(V,E)$ be an undirected graph with edge weights $p$. 
	Let $M$ be a matroid on edge set $E$ given by a polynomial time independence oracle.
	Let $x \in \PB{M}$.
	Let $\mathcal{L} \subseteq 2^V$ be a laminar family.
	Then, we can find in polynomial time an independent set $I \subseteq E$ with 
	\begin{itemize}
		\item $p(I) \ge \frac{63}{64}p^\top x$, and
		\item $|\delta(L) \cap I| \le \lceil 2x(\delta(L))\rceil+10$ for each $L \in \mathcal{L}$.
	\end{itemize} 
\end{lemma}

For $M$ being a refinement of the graphic matroid, \cref{lem:thin_trees_prize_preserving} yields a forest $F$ with $|\delta(L) \cap F| \le 2x(\delta(L))+11$ for each $L \in \mathcal{L}$.

\begin{lemma}\label{lem:thin_trees_face_preserving}
	Let $G=(V,E)$ be an undirected connected graph, and let $M$ be a refinement of the graphic matroid on edge set $E$.
	Let $x \in \PB{M}$.
	Let $\mathcal{L} \subseteq 2^V$ be a laminar family of nonempty proper subsets.
	Then, we can find in polynomial time a basis $B \subseteq E$ of $M$ with
	\begin{equation*}
		|\delta(L) \cap B| \le \tfrac{70}{3} x(\delta(L)) -\tfrac{34}{3} \qquad \forall \, L \in \mathcal{L}.
	\end{equation*}
\end{lemma}

We start by observing that, indeed, \cref{lem:thin_trees_prize_preserving,lem:thin_trees_face_preserving} imply \cref{thm:intro-laminar}.

\begin{proof}[Proof of \cref{thm:intro-laminar}]
Apply \cref{lem:matroid_equivalence2} to the cut constraints for the nonempty proper sets in $\mathcal L$. By \cref{lem:thin_trees_prize_preserving,lem:thin_trees_face_preserving}, the prize fraction is $\alpha=63/64$, the multiplicative violations of the cut constraints sum to $2+70/3=76/3$, and the additive terms sum to $11-34/3=-1/3$. Thus the cost-preserving tree satisfies
\[
|\delta(L)\cap T|\le \tfrac{64}{63}\cdot\tfrac{76}{3}\,x(\delta(L))-\tfrac13 < 26 x(\delta(L))
\]
for each nonempty proper cut $L \in \mathcal{L}$.
Since the cuts of $\emptyset$ and $V$ are empty, this proves \cref{thm:intro-laminar}.	
\end{proof}

In the following, we discuss how we obtain \cref{lem:thin_trees_prize_preserving,lem:thin_trees_face_preserving}.

\subsection{Approximately prize-preserving thin forests}

\cref{lem:thin_trees_prize_preserving} is the first of the two ingredients needed for \cref{lem:matroid_equivalence2}: it asks for an independent set that keeps a constant fraction of the prize $p^\top x$ while approximately satisfying the laminar cut constraints. We prove it for an arbitrary matroid on the edge set of the graph.

\paragraph{A simple randomized route.}
Before giving our argument, we sketch a quick randomized way to obtain such an independent set \emph{in expectation} with weaker constants. This is all that \cref{lem:matroid_equivalence2} requires from this step in order to yield a cost-preserving laminar-thin tree.
The tool is a \emph{contention resolution scheme} (CRS) \cite{CVZ14}. We first scale $x$ down by a constant $b\in(0,1)$ and sample a random set $R\subseteq E$ containing each $e\in E$ independently with probability $b\,x_e$. The CRS then discards edges of $R$ so that the remaining set $I$ is independent in the matroid and approximately satisfies the cut constraints.
CRSs can be designed separately for each constraint type and then combined. Here we combine one for matroids with one for laminar cut families. Strong CRSs exist for both~\cite{CVZ14}\footnote{Selecting edges subject to laminar cut constraints is a rephrasing of the unsplittable flow problem on trees studied in~\cite{CVZ14}.}, which directly gives a randomized algorithm returning the desired $I$ in expectation.

\paragraph{The stronger, deterministic route.}
To obtain the stronger guarantee of \cref{lem:thin_trees_prize_preserving}, we do not round all coordinates of $x$ independently, but instead use pipage rounding \cite{AS04} to directly obtain an independent set.
At each step, pipage rounding shifts weight from one fractional coordinate to another along an edge direction of the matroid base polytope. The process stays within the polytope and ends at the incidence vector of a basis.
The challenge is to bound the prize loss when making the resulting edge set thin. 
To do so, we use a set function as a proxy for the prize retained by a greedy selection rule we apply from \cite{CVZ14}.
The negative of its multilinear extension is concave under swaps, so deterministic pipage rounding rounds $x$ to a basis without decreasing the value of this multilinear extension.
Applying the greedy rule to this basis gives a thin independent set that retains at least $63/64$ of the fractional prize.
See \cref{sec:thin_trees} for full details.

\subsection{Face-preserving thin trees}

To obtain a face-preserving thin tree with respect to a point $x\in Q$ in the spanning tree polytope $Q$, it is useful to first replace $x$ by a more structured point $y\in F_Q(x)$, which is easier to round.
This is reminiscent of the approach by \textcite{KO23}, who construct a laminar-thin tree by first aligning the given point in the spanning tree polytope with the laminar cut family $\mathcal{L}$.
More precisely, they construct a laminar family $\overline{\mathcal{L}}$ and a point $y = \mathcal{O}(x)$ in the spanning tree polytope such that each set $\overline{L}\in \overline{\mathcal{L}}$ corresponds to a $y$-tight spanning tree constraint, and every spanning tree that is thin on $\overline{\mathcal L}$ with respect to $y$ is also thin on $\mathcal L$ with respect to $x$, up to a constant factor.
\cite{KO23} then show that this more structured setting allows for using an iterative relaxation approach to get a thin tree, because the alignment leads to higher sparsity of vertex solutions to the natural LP relaxation.
Unfortunately, their alignment procedure is not face-preserving, and therefore cannot be used in our setting to obtain a spanning tree without cost loss.

Instead, we only aim for what we call a \emph{weakly aligned} point, to be able to be face-preserving.
More precisely, we construct a point $y\in F_Q(x)$ with $y = \mathcal{O}(x)$, such that for each $L \in \mathcal{L}$, the total fractional mass of internal edges outside tight spanning tree constraints wholly contained in \(E[L]\) is bounded by \(\mathcal O(x(\delta(L)))\).
The following lemma shows that we can efficiently find a weakly-aligned point and also formalizes the weak alignment property through 
\Cref{eq:weak_alignment}.
\begin{lemma}[Weak alignment lemma]\label{lem:partial_alignment_new}
	Let $G=(V,E)$ be an undirected connected graph, and let $M$ be a refinement of the graphic matroid on edge set $E$.
	Let $\mathcal{L} \subseteq 2^V$ be a laminar family of nonempty proper subsets.
	For each point $x \in \PB{M}$ and each $\alpha > 1$, we can in polynomial time compute $k \in \mathbb{Z}_{\ge 0}$ and a refinement $N = \bigoplus_{i=1}^k N_i$ of $M$ and $y\in\PB{N}$ such that $y\le\alpha x$ and
	\begin{equation}\label{eq:weak_alignment}
		y(E[L] \setminus D_L)\le \frac{\alpha}{\alpha-1} \left( x(\delta(L)) -1 \right)
		\quad\forall\; L \in\mathcal L,
	\end{equation}
	where $D_L \coloneqq\bigcup_{i \in [k] :E_i\subseteq E[L]}E_i$, with $E_i\subseteq E$ being the ground set of $N_i$.
\end{lemma}

\begin{figure}[t]
	\centering
	\begin{adjustbox}{max width=\linewidth}
	\begin{tikzpicture}[font=\small, color=black,
		component/.style={draw=teal!55!black, fill=teal!9, thin},
		heavy/.style={teal!75!black, line width=1.5pt},
		remaining/.style={blue!65!black, thick, dashed},
		cut/.style={orange!80!black, thick}]
		\draw[black!65, fill=black!3] (0,0) ellipse [x radius=3.35cm, y radius=1.15cm];
		\node at (-3.04,.12) {$L$};
		\draw[component] (-1.60,0) ellipse [x radius=1.13cm, y radius=.66cm];
		\draw[component] (1.80,0) ellipse [x radius=.76cm, y radius=.44cm];
		\coordinate (a) at (-2.37,-.17); \coordinate (b) at (-1.90,.36);
		\coordinate (c) at (-.83,.08);   \coordinate (d) at (-1.46,-.42);
		\coordinate (e) at (1.28,.04);    \coordinate (f) at (1.91,.25);
		\coordinate (g) at (2.24,-.18);  \coordinate (s) at (0.5,-0.7);
		\coordinate (out1) at (2.5,1.60);
		\coordinate (out2) at (-2.0,-1.50);
		\coordinate (out3) at (.8,-1.7);
		\draw[cut] (out1)--(f) (out2)--(d) (out3)--(s);
		\draw[remaining] (c)--(e) (g)--(s);
		\draw[heavy] (a)--(b)--(c)--(d)--cycle (b)--(d) (e)--(f)--(g)--cycle;
		\foreach \v in {a,b,c,d,e,f,g,s,out1,out2,out3} \fill (\v) circle[radius=1.65pt];
		\draw[heavy]     (3.8,.63)--(4.5,.63) node[right,black] {$D_L$};
		\draw[remaining] (3.8,.06)--(4.5,.06) node[right,black] {$E[L]\setminus D_L$};
		\draw[cut]       (3.8,-.51)--(4.5,-.51) node[right,black] {$\delta(L)$};
	\end{tikzpicture}
	\end{adjustbox}
	\caption{Weak alignment for a set $L\in\mathcal L$.
	In this example, each vertex set shaded in teal defines a tight spanning-tree constraint at $y$, and the solid teal edges within these sets form $D_L$.
	The orange edges form the cut $\delta(L)$, and the dashed blue edges form $E[L]\setminus D_L$.
	To bound the cut load when rounding fractional solutions, we control a proxy that counts orange edges once and dashed edges twice.
	Weak alignment guarantees $y(E[L]\setminus D_L)\le\tfrac{\alpha}{\alpha-1}\bigl(x(\delta(L))-1\bigr)$, bounding the additional internal load in terms of the original fractional cut load.}
	\label{fig:partial-alignment}
\end{figure}
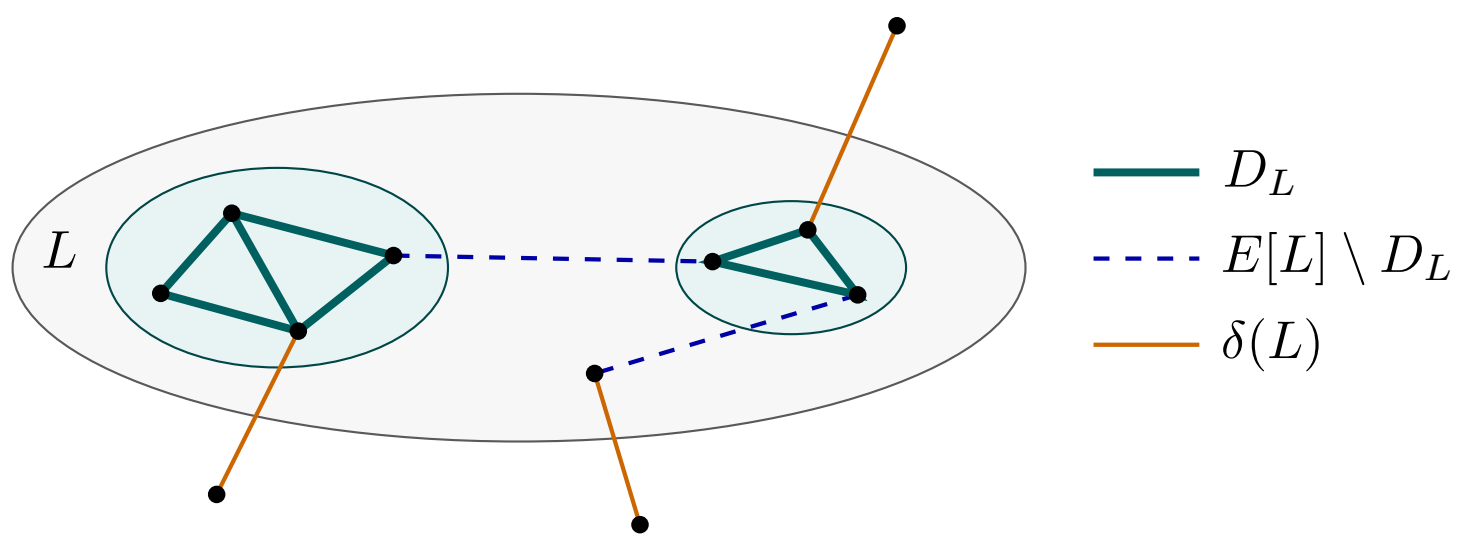

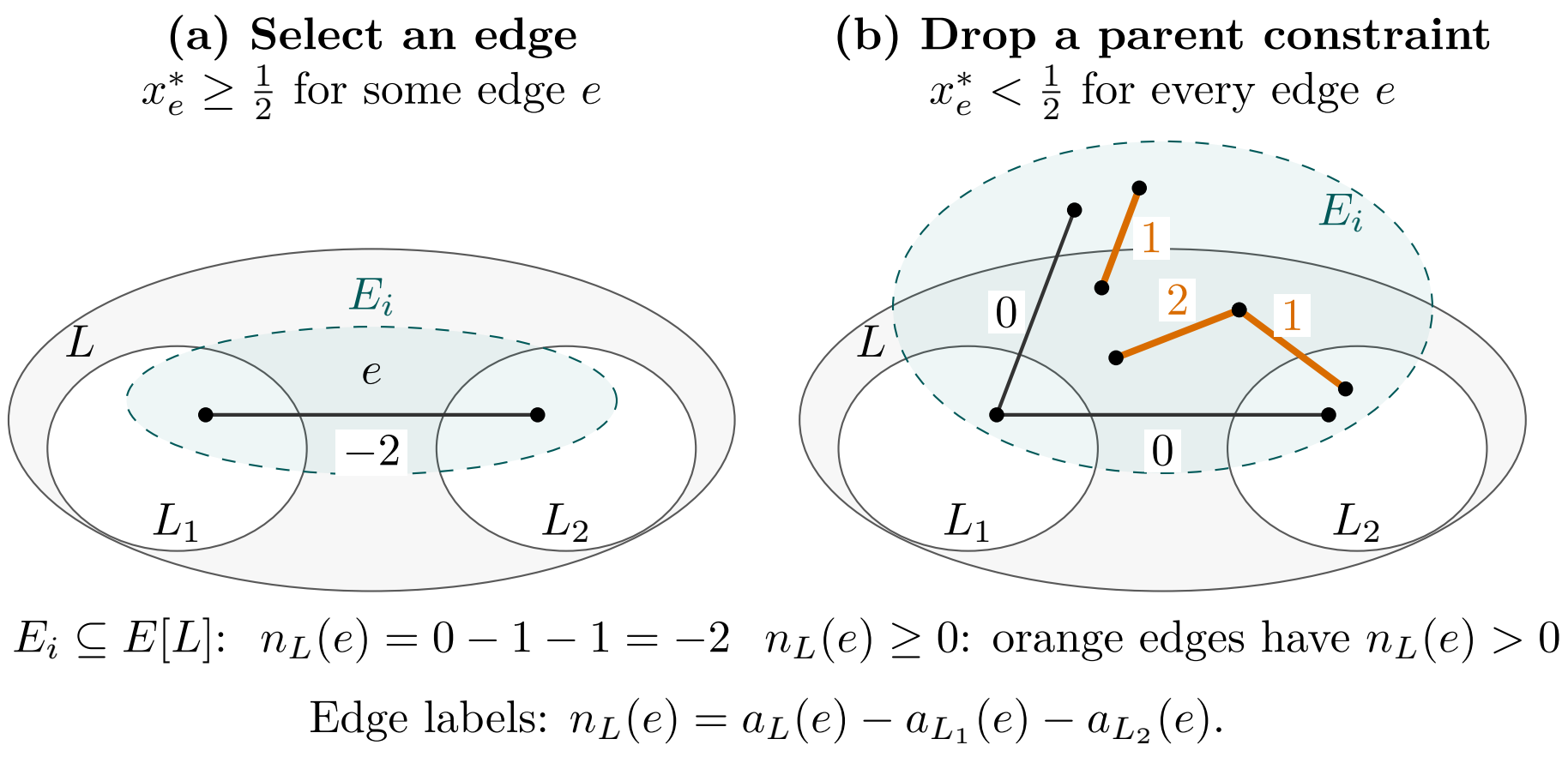
\begin{figure}[t]
	\centering
	\begin{tikzpicture}[font=\small, color=black,
		parent/.style={draw=black!65, fill=black!3},
		child/.style={draw=black!65, fill=white},
		summand/.style={draw=teal!70!black, dashed, fill=teal, fill opacity=.07},
		edge/.style={black!80, thick},
		extra/.style={orange!85!black, line width=1.5pt},
		coefficient/.style={fill=white, inner sep=1.5pt}]
		\foreach \offset/\panel in {0/1,6.1/2} {
			\begin{scope}[xshift=\offset cm]
				\draw[parent] (0,0) ellipse [x radius=2.8cm, y radius=1.32cm];
				\node at (-2.25,.61) {$L$};
				\foreach \pos/\i in {-1.5/1,1.5/2} {
					\draw[child] (\pos,-.22) ellipse [x radius=1cm, y radius=.79cm];
					\node at (\pos,-.79) {$L_{\i}$};
				}
				\ifnum\panel=1
					\node[font=\small\bfseries] at (0,2.95) {(a) Select an edge};
					\node at (0,2.53) {$x_e^*\ge\tfrac12$ for some edge $e$};
					\draw[summand] (0,.15) ellipse [x radius=1.89cm, y radius=.57cm];
					\node[teal!70!black] at (0,.95) {$E_i$};
					\node[above=3pt] at (0,.04) {$e$};
					\node[below=3pt,coefficient] at (0,.04) {$-2$};
					\node at (0,-1.70) {$E_i\subseteq E[L]$: $\;n_L(e)=0-1-1=-2$};
				\else
					\node[font=\small\bfseries] at (0,2.95) {(b) Drop a parent constraint};
					\node at (0,2.53) {$x_e^*<\tfrac12$ for every edge $e$};
					\draw[summand] (0,.87) ellipse [x radius=2.08cm, y radius=1.28cm];
					\node[teal!70!black] at (1.38,1.60) {$E_i$};
					\draw[edge] (-1.28,.04)-- node[midway,left=2pt,coefficient,text=black] {$0$} (-.68,1.62);
					\fill (-.68,1.62) circle[radius=1.65pt];
					\foreach \xa/\ya/\xb/\yb/\diff/\side in {
						-.47/1.02/-.18/1.79/1/right,
						-.36/.48/.59/.85/2/above,
						.59/.85/1.41/.24/1/above} {
						\draw[extra] (\xa,\ya)-- node[midway,\side=2pt,coefficient,text=orange!85!black] {$\diff$} (\xb,\yb);
						\fill (\xa,\ya) circle[radius=1.65pt];
						\fill (\xb,\yb) circle[radius=1.65pt];
					}
					\node[below=3pt,coefficient] at (0,.04) {$0$};
					\node at (0,-1.70) {$n_L(e)\ge0$: orange edges have $n_L(e)>0$};
				\fi
				\draw[edge] (-1.28,.04)--(1.28,.04);
				\foreach \x in {-1.28,1.28} \fill (\x,.04) circle[radius=1.65pt];
			\end{scope}
		}
		\node at (3.05,-2.32) {Edge labels: $n_L(e)=a_L(e)-a_{L_1}(e)-a_{L_2}(e)$.};
	\end{tikzpicture}
	\caption{Two schematic configurations in the iterative relaxation of \cref{lem:rounding_thin_trees}.
	The coefficient $a_S(e)$ counts a cut edge once, an internal edge outside $D_S$ twice, and all other edges zero times.
	We obtain the initial bound $b_S$ by rounding the fractional load $a_S(y)=\sum_{e\in E}a_S(e)y_e$ up to the nearest multiple of $\tfrac12$. During the algorithm, we may decrease these bounds, always keeping them half-integral.
	The inductive goal is to find a basis $B$ satisfying $a_S(B)\le4b_S$ for every constraint $S$.
	Dashed teal ovals indicate the original ground sets of matroid summands; edge labels show the parent's coefficient minus the sum of its children's coefficients.
	\textbf{(a)} At the current extreme point $x^*$, an edge with $x_e^*\ge\tfrac12$ can be selected and contracted, reducing each bound whose constraint counts it by $\tfrac12$.
	The displayed edge is counted by both children but not by the parent, giving $n_L(e)=-2$.
	\textbf{(b)} If every $x_e^*<\tfrac12$, the counting argument in \cref{claim:relax} guarantees a tight parent constraint $L$, illustrated with two tight children, for which $n_L(e)\ge0$ for every edge and the average of the values $x_e^*$, weighted by $n_L(e)$, is at least $\tfrac14$.
	It yields the first inequality in
	$\sum_{e\in E}n_L(e)\le4\sum_{e\in E}n_L(e)x_e^*=4\bigl(b_L-b_{L_1}-b_{L_2}\bigr)$,
	where the equality follows from tightness.
	Thus, even selecting all orange edges adds at most $4(b_L-b_{L_1}-b_{L_2})$ to the load already counted by the children.
	Together with the children's bounds, this yields $a_L(B)\le4b_{L_1}+4b_{L_2}+4(b_L-b_{L_1}-b_{L_2})=4b_L$, so the parent constraint can be dropped.}
	\label{fig:parent-relaxation}
\end{figure}

A crucial property of weak alignment is that it allows us to use a well-structured proxy to bound the number of edges chosen in $\delta(L)$.
The proxy counts each chosen edge in $\delta(L)$ once and each chosen edge in $E[L]\setminus D_L$ twice.
See \cref{fig:partial-alignment} for an illustration.
This proxy allows us, similar to \cite{KO23}, to use an iterative relaxation algorithm to obtain a desired laminar-thin tree.
More precisely, in each iteration, we either round up an edge of value at least $\frac12$, or we can identify a constraint to relax. 
\cref{fig:parent-relaxation} sketches the two cases of the iterative relaxation.
(See \cref{sec:thin_trees} for a formal description of the algorithm and a correctness proof.)
This leads to the following result.

\begin{lemma}\label{lem:rounding_thin_trees}
	Let $G=(V,E)$ be an undirected graph, and let $\mathcal{L} \subseteq 2^V$ be a laminar family.
	Let $k \in \mathbb{Z}_{\ge 0}$, and let $M=\bigoplus_{i=1}^k M_i$ be a direct sum of matroids $M_i$, with pairwise disjoint nonempty ground sets $E_i$ and total ground set $E=\bigcup_{i=1}^k E_i$.
	For each $y \in \PB{M}$ we can find in polynomial time a basis $B$ of $M$ with 
	\begin{equation*}
		|\delta(L) \cap B| \le 4 y(\delta(L)) + 8 y(E[L] \setminus D_L) + 2  
		\quad\forall\; L \in\mathcal L,
	\end{equation*}
	where $D_L \coloneqq\bigcup_{i \in [k]:E_i\subseteq E[L]}E_i$.
\end{lemma}

Indeed, \cref{lem:partial_alignment_new,lem:rounding_thin_trees} directly yield \cref{lem:thin_trees_face_preserving}.

\begin{proof}[Proof of \cref{lem:thin_trees_face_preserving}]
	Apply \cref{lem:partial_alignment_new} with $\alpha=5/2$ to $\mathcal L\setminus\{\emptyset,V\}$, and then apply \cref{lem:rounding_thin_trees} to the resulting refinement $N$ and point $y$ with the same laminar family. The resulting basis $B$ is also a basis of $M$, since $N$ refines $M$. For each $L\in\mathcal L\setminus\{\emptyset,V\}$, the two lemmas give
	\[
	\begin{aligned}
		|\delta(L)\cap B|
		&\le 4y(\delta(L))+8y(E[L]\setminus D_L)+2 \\
		&\le 10x(\delta(L))+\tfrac{40}{3}\bigl(x(\delta(L))-1\bigr)+2 \\
		&=\tfrac{70}{3}x(\delta(L))-\tfrac{34}{3}.
	\end{aligned}
	\]
	Both constructions run in polynomial time.
\end{proof}
\ifbool{ipcosubm}{\section{Cost-preserving rounding for general polytopes}
\label{sec:equivalence}
\label{sec:general_polytopes}

In this section, we prove \cref{lem:general_polytopes}.
}{\section{A characterization for cost-preserving rounding}\label{sec:equivalence}

In this section, we prove \cref{lem:general_polytopes,lem:matroid_equivalence2}.

\subsection{Cost-preserving rounding for general polytopes}
\label{sec:general_polytopes}

We start with showing \cref{lem:general_polytopes}.
}To show that \ref{item:unweighted} implies \ref{item:weighted}, we construct a linear objective $w=c+p$. Here $c$ is the given cost vector, and \(p\) is a nonnegative prize vector obtained from the optimal dual multipliers after relaxing the constraint bounds by a factor of \(1/\alpha\). 
Any optimal solution \(z\) of the relaxed primal then minimizes \(w\) over \(P\) and collects $1 / \alpha$ times the prize of $x$.
Since $w$ is constant on the minimal face of $z$, rounding within this face keeps the sum of cost and prize at its minimum. It therefore suffices for the rounded point to retain an $\alpha$-fraction of the prize of $z$. Its prize is then at least that of $x$, while its sum of cost and prize is at most that of $x$, so its cost is at most $c^\top x$.

For the converse, we choose a cost vector with two terms. The first is constant on $F_P(x)$ and gives every integral point outside this face a penalty larger than any possible prize gain over $x$. The second term is the negative of the given prize vector. A rounded point of cost at most $c^\top x$ must therefore lie in $F_P(x)$, where the cost guarantee implies that its prize is at least $p^\top x$.

\begin{proof}[Proof of \cref{lem:general_polytopes}]
We start with showing that \ref{item:unweighted} implies \ref{item:weighted}.
Assume that there are constants $\alpha \in (0,1]$ and $\beta > 0$ such that for each $x \in P$ and prizes $p \in \mathbb{R}_{\ge 0}^n$, we can find an integral point $y \in F_P(x)$ with
\begin{equation*}
	p^\top y \ge \alpha p^\top x,
	\quad \text{and} \quad
	Ay \le \beta Ax +d\one.
\end{equation*}
Let $c \in \mathbb{R}^n$.
Consider the relaxed linear program
\begin{equation}\label{eq:primal}
	\min\bigl\{\ip{c}{z}:z\in P,\ Az\le \alpha^{-1} Ax\bigr\}.
\end{equation}
Its feasible region is nonempty, since it contains $x$ as $A,x\ge0$ and $\alpha\in (0,1]$.
Dualizing only the constraints \(Az\le\alpha^{-1}Ax\) gives the Lagrangian dual
\begin{equation}\label{eq:dual}
	\max_{\mu\in\mathbb{R}_{\ge0}^{m}}
	\left\{\min_{z\in P}(c+A^{\mathsf T}\mu)^{\mathsf T}z
	-\alpha^{-1} \mu^{\mathsf T}Ax\right\}.
\end{equation}
Linear programming duality guarantees an optimal multiplier $\mu^*\ge0$ and equality of the primal and dual optima.
Set $p\coloneqq A^{\mathsf T}\mu^*$ and $w\coloneqq c+p$.
Let $z^*$ be an optimal solution of \eqref{eq:primal}.
By strong Lagrangian duality, $z^*$ attains the inner minimum in \eqref{eq:dual} at $\mu^*$, so it minimizes $w$ over $P$.
Separately, complementary slackness for the constraints \(Az\le\alpha^{-1}Ax\) gives $\mu_i^*((Az^*)_i-\alpha^{-1}(Ax)_i)=0$ for every $i \in [m]$.
Summing these equalities and using $p=A^{\mathsf T}\mu^*$ gives $\alpha \ip{p}{z^*}=\ip{p}{x}$.
Thus,
\begin{equation*}
	z^*\in\arg\min_{z\in P}\ip{w}{z},
	\quad \text{and} \quad
	\alpha \ip{p}{z^*}=\ip{p}{x}.
\end{equation*}
Since $A$ and $\mu^*$ are nonnegative, so is $p$.
By applying the face-preserving rounding algorithm to $z^*$, we can find in polynomial time an integral point $y \in F_P(z^*)$ with $p^\top y \ge \alpha p^\top z^*$, and $Ay \le \beta Az^* +d\one \le \frac{\beta}{\alpha} Ax +d\one$.
Note that the linear functional $w$ is constant on $F_P(z^*)$.
Consequently,
\begin{align*}
	c^\top y&=\ip{w}{z^*}-p^\top y
	\le\ip{w}{z^*}-\alpha\ip{p}{z^*}
	=\ip{w}{z^*}-\ip{p}{x}
	\le\ip{w}{x}-\ip{p}{x}
	=\ip{c}{x}.
\end{align*}
The optimal solutions $z^*$ and $\mu^*$ can be computed in polynomial time by solving \eqref{eq:primal} and \eqref{eq:dual} with the ellipsoid method.

Conversely, to show that \ref{item:weighted} implies \ref{item:unweighted}, suppose that there is a constant $\gamma>0$ such that
for every $x\in P$ and every cost vector $c \in \mathbb{R}^n$, we can find in polynomial time an integral point $y\in P$ with
\[
c^\top y\le c^\top x,
\quad \text{and} \quad
Ay\le \gamma Ax +d\one.
\]
Fix $x\in P$ and prizes $p \in \mathbb{R}_{\ge 0}^n$.
If $F_P(x)=P$, applying the cost-preserving algorithm with $c=-p$ yields the desired conclusion.
Otherwise, by the proof of Theorem 6.5.8 in \cite{GLS1988}, we can compute a supporting hyperplane $\{z\in\mathbb R^n:a^\top z=b\}$ with $F_P(x)=P\cap\{z\in\mathbb R^n:a^\top z=b\}$ and $P \subseteq \{z\in\mathbb R^n:a^\top z\ge b\}$ in polynomial time, where, after scaling, $a\in\mathbb Z^n$ and $b \in \mathbb{Z}$.

Compute $p^*:=\max_{z\in P}p^\top z$, and let $M:=1+p^*-p^\top x>0$.
Apply the cost-preserving algorithm at $x$ with cost vector
$c:=Ma-p$.
Its integral output $y\in P$ satisfies
\begin{equation}\label{eq:minimizer}
Ma^\top y - p^* \le (Ma^\top - p^\top) y \le (M a^\top - p^\top) x = M b - p^\top x,
\end{equation}
yielding
\[
M(a^\top y-b)
\le p^*-p^\top x
<M.
\]
Because $a^\top y-b$ is a nonnegative integer,
this implies $a^\top y=b$, and hence $y\in F_P(x)$.
Using $a^\top y=a^\top x = b$, the second inequality in \eqref{eq:minimizer} yields
\[
p^\top y\ge p^\top x.
\]
Moreover, $Ay\le \gamma Ax +d\one$ as guaranteed by the cost-preserving algorithm.
\end{proof}

\ifbool{ipcosubm}{\section*{AI acknowledgement}
We used ChatGPT-6 for assisting with technical statements and proofs, and also to improve the write-up.
\printbibliography
\clearpage
\appendix
}{}

\ifbool{ipcosubm}{\section{Cost-preserving rounding algorithms for matroids}
}{\subsection{Cost-preserving rounding algorithms for matroids}
}
\label{sec:general_matroids}

For matroid base polytopes, we can collect a constant fraction of the prize before committing to a basis. One algorithm finds an independent set $I$ that collects this prize, while a second finds a basis $B$. By matroid augmentation, we can extend $I$ to a basis using only elements of $B$. 
As the prizes are nonnegative, this preserves the prize collected by $I$ and bounds the load of the resulting basis by the sum of the loads of $I$ and $B$. To ensure that the resulting basis lies in the minimal face containing $x$, we run both algorithms on the refinement whose base polytope is that face. This is why \cref{lem:matroid_equivalence2} requires both algorithms to work for any given refinement.

\begin{proof}[Proof of \cref{lem:matroid_equivalence2}]
	The independence oracle gives polynomial-time optimization over $\PB{M}$ by the greedy algorithm, and hence polynomial-time separation.
	
	First, assume \ref{item:unweighted_matroid2}, and fix $x\in\PB{M}$ and prizes $p \in \mathbb{R}_{\ge 0}^N$.
	By \cref{lem:refinement}, we can efficiently construct a polynomial-time independence oracle for the matroid $M_x$ on $N$ with $\PB{M_x}=F_{\PB{M}}(x)$.
	
	Apply the two algorithms to the refinement $M_x$ and $x$, using the prize vector $p$ for the first algorithm, and let $I$ and $B$ be the independent set and basis returned, respectively.
	By their guarantees, $I$ is independent in $M_x$ and $B$ is a basis of $M_x$.
	Matroid augmentation therefore allows us to extend $I$ in polynomial time to a basis $B'$ of $M_x$ using only elements of $B$, so that $I\subseteq B'\subseteq I\cup B$.
	Since $p$ and $A$ are nonnegative, we obtain
	\[
	p(B')\ge p(I)\ge\alpha p^\top x,\quad \text{and} \quad
	A\chi^{B'}\le A\chi^I+A\chi^B\le(\beta_1+\beta_2)Ax+(d_1+d_2)\one.
	\]
	Moreover, $\chi^{B'}\in F_{\PB{M}}(x)$.
	Thus \cref{lem:general_polytopes}~\ref{item:unweighted} holds for $P=\PB{M}$ with $\beta=\beta_1+\beta_2$ and $d=d_1+d_2$.
	The equivalence of \cref{lem:general_polytopes}~\ref{item:unweighted} and \cref{lem:general_polytopes}~\ref{item:weighted} then yields \ref{item:weighted_matroid2} with $\gamma=(\beta_1+\beta_2)/\alpha$ and the same $d$.

	Conversely, assume \ref{item:weighted_matroid2}.
	Given a refinement $M'$ of $M$, $x\in\PB{M'}$, and prizes $p \in \mathbb{R}_{\ge 0}^N$, apply \cref{lem:general_polytopes} to $\PB{M}$ to find in polynomial time a basis $B$ in $F_{\PB{M}}(x)$ with
	\[
	p(B)\ge p^\top x,\quad \text{and} \quad A\chi^B\le\gamma Ax+d\one.
	\]
	Since $\PB{M'}$ is a face of $\PB{M}$ containing $x$, we have $F_{\PB{M}}(x)\subseteq\PB{M'}$.
	Thus $B$ is a basis of $M'$.
	Taking $I=B$ gives the first algorithm in \ref{item:unweighted_matroid2} with $\alpha=1$, $\beta_1=\gamma$, and $d_1=d$.
	Applying the same algorithm with $p=0$ gives the second algorithm in \ref{item:unweighted_matroid2} with $\beta_2=\gamma$ and $d_2=d$.
\end{proof}
\section{Laminar thin trees}
\label{sec:thin_trees}

We apply \cref{lem:matroid_equivalence2} to round a fractional spanning tree to a laminar-thin spanning tree at no cost loss. As discussed in \cref{sec:approach_overview}, this requires two algorithms for a given refinement of the graphic matroid. The first finds a thin forest that collects a constant fraction of the fractional prize (cf. \cref{lem:thin_trees_prize_preserving}). The second finds a thin spanning tree whose edges can be used to extend the forest to a basis of the refinement (cf. \cref{lem:thin_trees_face_preserving}).
We start with proving \cref{lem:thin_trees_prize_preserving}.

\subsection{Prize-preserving thin forests}

\begin{proof}[Proof of \cref{lem:thin_trees_prize_preserving}]
	We may assume $p\ge0$: replace negative prizes by zero and remove the corresponding edges from the returned set. Add $V$ and all singletons to $\mathcal L$ and  discard the empty set.
	
	We consider the following auxiliary graph. Form a tree $H$ rooted at $V$ by making each other set a child of its smallest strict superset. Each edge $e=\{v,w\}$ of $G$ gives a path $P_e$ between the leaves $\{v\}$ and $\{w\}$. The tree edge $h$ joining $L$ to its parent lies on $P_e$ exactly when $e\in\delta(L)$. Give $h$ capacity $u_h=\lceil x(\delta(L))\rceil$. We aim to find an independent set of profitable paths respecting these capacities. We may delete edges $e$ with $x_e=0$ from $M$, so every tree edge on a remaining path has positive capacity.

	We first give a simple randomized algorithm with a worse constant. 
	Set $b=1/32$ and form $R\subseteq E$ by including each edge $e$ independently with probability $b x_e$. Apply the contention resolution schemes of \textcite[Corollary~4.11 and Lemma~4.19]{CVZ14} for the matroid and for unit-demand paths in a tree to the same set $R$. Let $I$ be the intersection of their outputs. Then $I$ is independent in $M$ and respects all capacities. Conditional on $e\in R$, and using the guarantees in \cite{CVZ14}, the contention resolution scheme for the matroid keeps $e$ with probability at least $(1-\exp(-b))/b\ge 1-b$, and the contention resolution scheme for the unit-demand paths in a tree rejects $e$ with probability at most $2\mathrm e b/(1-\mathrm e b)$.
	A union bound then gives
	\[
	\Pr[e\in I] \ge bx_e \left( 1- b - \frac{2\mathrm e b}{1-\mathrm e b} \right) \ge 3b x_e/4.
	\]
	Thus $\mathbb E[p(I)]\ge \tfrac{3}{128} p^\top x$.

	We now give a deterministic algorithm based on deterministic pipage rounding. For each tree edge $h$ connecting $L \in \mathcal{L} \setminus \{V\}$ with its parent in $H$, increase its capacity to $u_h=\lceil 2 x(\delta(L)) \rceil+10$. Given a set $S$ of paths, process them in order of increasing depth of their highest vertex, breaking ties by an arbitrary but fixed order. 
	According to the greedy rule in \cite{CVZ14}, accept a path if adding it respects all capacities. We first bound the prize lost by this rule, and then use this bound to choose an independent set $S$ deterministically.
	
	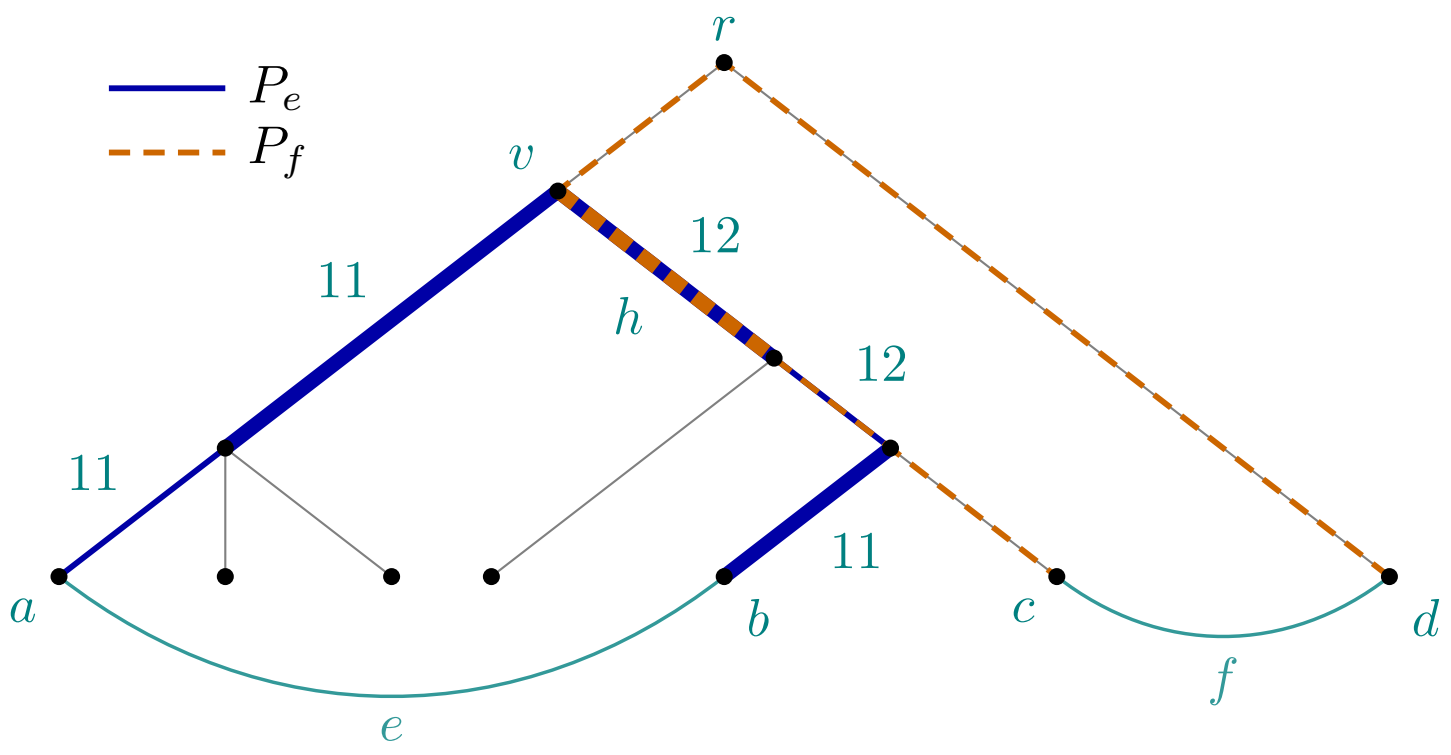
\begin{figure}[t]
		\centering
		\AI{\begin{adjustbox}{max width=\linewidth}
		\begin{tikzpicture}[
			x=1.1cm, y=.85cm, font=\small, line width=1.1pt, line join=round,
			path e/.style={blue!65!black},
			path f/.style={orange!80!black, dash pattern=on 4pt off 2.5pt},
			marked/.style={line width=2.8pt},
			link/.style={blue!50!green!80!white, line width=.65pt}
			]
			\coordinate[label=above:{$r$}]      (r) at (0,4);
			\coordinate[label=above left:{$v$}] (v) at (-1,3);
			\coordinate (m) at (-3,1);
			\coordinate (mcenter) at (-3,0);
			\coordinate (mright) at (-2,0);
			\coordinate (w) at (.3,1.7);
			\coordinate (t) at (1,1);
			\coordinate[label=below left:{$a$}]  (a) at (-4,0);
			\coordinate[label=below right:{$b$}] (b) at (0,0);
			\coordinate[label=below left:{$c$}]  (c) at (2,0);
			\coordinate[label=below right:{$d$}] (d) at (4,0);
			\coordinate (s) at (-1.4,0);

			\draw[link] (a) to[bend right=38] node[midway,below] {$e$} (b);
			\draw[link] (c) to[bend right=38] node[midway,below] {$f$} (d);

			\draw[gray,thin] (mcenter)--(m)--(mright) (w)--(s) (d)--(r)--(v) (t)--(c);
			\draw[path e,marked] (m)--(v)--(w) (t)--(b);
			\draw[path e] (a)--(m) (w)--(t);
			\draw[path f] (d)--(r)--(v) (w)--(t)--(c);
			\draw[path f,marked] (v)--(w);

			\path (v)--(m) node[midway,above left]  {$11$};
			\path (m)--(a) node[midway,above left]  {$11$};
			\path (v)--(w) node[midway,above right] {$12$};
			\path (v)--(w) node[midway,below left] {$h$};
			\path (w)--(t) node[midway,above right] {$12$};
			\path (t)--(b) node[midway,below right] {$11$};

			\foreach \u in {r,v,m,mcenter,mright,w,t,a,b,c,d,s}
				\fill[black] (\u) circle[radius=1.6pt];

			\draw[path e] (-3.7,3.8)--(-3,3.8) node[right,black] {$P_e$};
			\draw[path f] (-3.7,3.3)--(-3,3.3) node[right,black] {$P_f$};
		\end{tikzpicture}
		\end{adjustbox}}
		\caption{For edges $e = \{a,b\}$ and $f = \{c,d\}$ in $E$, the figure shows the corresponding paths $P_e$ and $P_f$ in $H$.
		Assume $x_e = x_f = \frac12$. 
		The numbers next to the edges show the capacity for tree edges in $P_e$ for the deterministic algorithm. 
		Edges in $H_e$ are marked bold. 
		The highest vertex $r$ of $P_f$ lies above the highest vertex $v$ of $P_e$, so $P_f$ precedes $P_e$ in the fixed ordering. 
		Since $P_f$ uses the marked edge $h$, we have $S_{e,h}=\{f\}$ in this two-path example.}
		\label{fig:marked-path-edges}
	\end{figure}

	Fix a path $P_e$ and consider one of the two parts from its highest vertex to a leaf. Walking down this part, mark the first tree edge and thereafter mark an edge whenever its capacity is strictly smaller than all preceding capacities on that part. Let $H_e$ be the set of edges marked on the two parts. For $h\in H_e$, let $S_{e,h}$ consist of all \(f\in E\) such that \(P_f\) precedes \(P_e\) and uses \(h\).
	By the ordering of the paths, the load from earlier accepted paths is nonincreasing along the part. 
	See \cref{fig:marked-path-edges} for an illustration.
	If the capacity of an edge is exhausted when $P_e$ is considered, the capacity of the last marked edge at or above it is also exhausted: its capacity is no larger and its load is at least as large. Hence rejection of $e\in S$ implies $|S\cap S_{e,h}|\ge u_h$ for some $h\in H_e$.

	Fix a set $S$ of paths, and let $I(S)$ be the subset accepted by the greedy rule. If $e\in S$ is rejected, then, by the above discussion, for at least one marked edge $h\in H_e$ we have $2^{|S\cap S_{e,h}|-u_h}\ge 1$. Thus,
	\begin{equation}\label{eq:proxy}
		p(I(S))\ge \phi(S)\coloneqq
		\sum_{e\in S}p_e\left(1-\sum_{h\in H_e}2^{|S\cap S_{e,h}|-u_h}\right).
	\end{equation}
	We now consider the multilinear extension $\Phi$ of $\phi$. For $z\in[0,1]^E$, define
	\[
		\Phi(z)=\mathbb{E}[\phi(S_z)],
	\]
	where $S_z$ includes each $e\in E$ independently with probability $z_e$.
	To compute $\Phi(z)$, consider the contribution of a single edge $e$ to $\phi(S_z)$. It contributes only if it is sampled, which happens with probability $z_e$. For each $h\in H_e$, the term $2^{|S_z\cap S_{e,h}|}$ has one factor for each $f\in S_{e,h}$. 
	This factor is $2$ if $f$ is sampled and $1$ otherwise, so its expectation is $2z_f+(1-z_f)=1+z_f$. 
	The factors are independent, and since $e\notin S_{e,h}$, they are also independent of whether $e$ is sampled. 
	Summing the expected contributions of all edges gives
	\[
		\Phi(z)=\sum_{e\in E}p_e z_e
		\left(1-\sum_{h\in H_e}2^{-u_h}\prod_{f\in S_{e,h}}(1+z_f)\right)
	\]
	for each $z\in[0,1]^E$.
	It remains to show that this proxy is large at $x$ and that we can round $x$ to a basis $B$ without decreasing it.
	Using \eqref{eq:proxy}, this then yields $p(I(B)) \ge \phi(B) = \Phi(\chi^B) \ge \Phi(x)$.

	For brevity, let $q_h=x(\delta(L))$ be the fractional load on the tree edge $h$ joining $L$ to its parent. For each $h\in H_e$, we have $x(S_{e,h})\le q_h$ and $2q_h\le u_h-10$. Hence
	\[
		2^{-u_h}\prod_{f\in S_{e,h}}(1+x_f)
		\le 2^{-u_h}e^{q_h}
		\le 2^{-10}\left(\tfrac{\sqrt{\mathrm e}}{2}\right)^{u_h-10}
		\le 2^{-10}\left(\tfrac78\right)^{u_h-10}.
	\]
	On each of the two parts of $P_e$, the marked values $u_h-10$ are distinct positive integers, since $q_h\ge x_e>0$ on $P_e$. Thus,
	\[
	\sum_{h \in H_e}  2^{-10}\left(\tfrac78\right)^{u_h-10} \le 2\cdot 2^{-10}\sum_{j=1}^\infty \left(\tfrac78\right)^j=14/1024<1/64,
	\] 
	and $\Phi(x)\ge \tfrac{63}{64}p^\top x$.

	To round $x$, we use deterministic pipage rounding in the formulation of \textcite[Theorem~3.2]{HO14}. Given $x\in\PB{M}$ and a value oracle for a function $g:\PB{M}\to\mathbb R$ that is concave under swaps, it finds in polynomial time a basis $S$ of $M$ with $g(\chi^S)\le g(x)$. Here, concavity under swaps means that replacing two coordinates $z_i,z_j$ by $z_i+t,z_j-t$ gives a concave function of $t$ on the feasible interval.
	The idea is that if this function is concave, then rounding in one of the two directions will not increase the objective value.

	Take $g=-\Phi$, which we can evaluate in polynomial time from the displayed products. Since $e\notin S_{e,h}$, $g$ is multilinear, and all its nonlinear monomials have nonnegative coefficients. Along a swap, each term is affine in $t$ or a nonnegative multiple of $(z_i+t)(z_j-t)$. Thus $g$ is concave under swaps, and pipage rounding gives a basis $S$ with $\Phi(\chi^S)\ge\Phi(x)$. Applying the greedy rule to $S$ gives an independent set $I(S) \subseteq S$ that respects all capacities and satisfies
	\[
	p(I(S))\ge\phi(S)=\Phi(\chi^S)\ge \Phi(x) \ge \tfrac{63}{64}p^\top x,
	\]
	as desired.
\end{proof}

\subsection{Face-preserving thin trees}

As discussed in \cref{sec:approach_overview}, we obtain a thin tree in the prescribed face in two steps. We first weakly align the fractional point with the laminar family (\cref{lem:partial_alignment_new}) and then round it to a tree in the resulting refinement (\cref{lem:rounding_thin_trees}). Together, these steps prove \cref{lem:thin_trees_face_preserving}.

\subsubsection{Weak alignment with laminar families}

We prove the weak alignment step for an arbitrary matroid and a laminar family of subsets of its ground set. In the graphic setting, these subsets are the edge sets induced by the members of $\mathcal L$.

\begin{lemma}\label{lem:partial_alignment_matroid}
	Let $M$ be a matroid on ground set $E$ given by an independence oracle, and let $\mathcal R\subseteq 2^E$ be a laminar family.
	For each $x\in\PB{M}$ and each $\alpha>1$, we can compute in polynomial time an integer $k\ge 0$, a refinement $N=\bigoplus_{i=1}^k N_i$ of $M$, and a point $y\in\PB{N}$ such that $y\le\alpha x$ and
	\[
		y(D\setminus U_D)\le \frac{\alpha}{\alpha-1}\bigl(r_M(D)-x(D)\bigr)
		\qquad\forall D\in\mathcal R,
	\]
	where $U_D=\bigcup_{i \in [k]:E_i\subseteq D}E_i$ with $E_i$ being the ground set of $N_i$.
\end{lemma}

\begin{proof}[Proof of \cref{lem:partial_alignment_new}]
	Apply \cref{lem:partial_alignment_matroid} to the laminar family $\mathcal R=\{E[L]:L\in\mathcal L\}$. For $D=E[L]$, we have $U_D=D_L$, so it suffices to show $r_M(E[L])-x(E[L])\le x(\delta(L))-1$. 
	
	Since $M$ refines the graphic matroid and $L$ is nonempty and proper, we have $r_M(E[L])\le |L|-1$, $x(E)=|V|-1$, and $x(E[V\setminus L])\le |V\setminus L|-1$. Thus
	\[
	x(E[L])=x(E)-x(E[V\setminus L])-x(\delta(L))\ge |L|-x(\delta(L)) \ge r_M(E[L])+1-x(\delta(L)),
	\]
	as desired.
\end{proof}

To prove \cref{lem:partial_alignment_matroid}, we identify, within each $D\in\mathcal R$, the unique largest set $A(D)$ whose rank constraint can be made tight by scaling with factor $\alpha$. We use these sets to construct a refinement and show that the resulting point places little weight on $D\setminus A(D)$ whenever $x(D)$ is close to $r_M(D)$.

For $A\subseteq E$, the restriction $M{|A}$ denotes the matroid on $A$ whose independent sets are the independent sets of $M$ contained in $A$. The contraction $M/A$ is the matroid on $E\setminus A$ whose rank function is given by $r_{M/A}(T)=r_M(A\cup T)-r_M(A)$ for $T\subseteq E\setminus A$.
In words, for a basis $B_A$ of $M{|A}$, $I$ is independent in $M / A$ if and only if $I \cup B_A$ is independent in $M$.

\begin{proof}[Proof of \cref{lem:partial_alignment_matroid}]
	Add $E$ to $\mathcal R$. We say that $A\subseteq E$ can be made tight if there is a point $w\in\PB{M{|A}}$ with $w\le \alpha x|_{A}$.

	\begin{claim}\label{claim:largest_tight_subset}
		For each $D\in\mathcal R$, we can find in polynomial time the unique largest subset $A(D)\subseteq D$ that can be made tight.
		Moreover, this set satisfies
		$\alpha x(D\setminus A(D))\le r_M(D)-r_M(A(D))$.
	\end{claim}

	If $D\subseteq D'$, then $A(D)$ is also a candidate in $D'$, so $A(D)\subseteq A(D')$. Moreover, $A(E)=E$ because $x\in\PB{M}$.

	Next, we construct the refinement $N$.
	Order the sets $D_1,\ldots,D_k$ of $\mathcal R$ such that $D_i \subseteq D_j$ implies $i \le j$. 
	Let $C_0=\emptyset$, and $C_i=\bigcup_{j\le i}A(D_j)$ for $i \in [k]$. Since $C_k=E$, the matroid $N=\bigoplus_{i=1}^k N_i$ with $N_i=(M|C_i)/C_{i-1}$ on ground set $E_i = C_i \setminus C_{i-1}$ is a refinement of $M$; its bases are exactly the bases of $M$ that are tight on every $C_i$.
	
	\begin{claim}\label{claim:bounded_refinement_point}
		There is a point $y\in\PB{N}$ with $y\le \alpha x$.
	\end{claim}
	
	Before proving the two claims, we show that every such $y$ satisfies the desired inequalities.
	Note that such a point $y$ can be computed in polynomial time.

	Fix $D=D_i\in\mathcal R$ and write $A=A(D)$. By \cref{claim:largest_tight_subset} and $x(A)\le r_M(A)$, we have
	\[
		r_M(D)-x(D)
		\ge r_M(D)-r_M(A)-x(D\setminus A)
		\ge (\alpha-1)x(D\setminus A).
	\]
	Moreover, $A\subseteq U_D$. Indeed, each element of $A$ belongs to some $E_j$ with $j\le i$. Since $E_j\subseteq D_j$ and $D_j$ meets $D$, laminarity and the ordering give $E_j\subseteq D_j\subseteq D$.
	Thus every $y$ as in \cref{claim:bounded_refinement_point} satisfies
	\[
		y(D\setminus U_D)\le  y(D\setminus A) \le \alpha x(D\setminus A)
		\le \frac{\alpha}{\alpha-1}\bigl(r_M(D)-x(D)\bigr),
	\]
	as desired.
	It remains to prove \cref{claim:largest_tight_subset,claim:bounded_refinement_point}.

	\begin{proof}[Proof of \cref{claim:largest_tight_subset}]
		Fix $D\in\mathcal R$. In the independence polytope $\PI{M|D}$ of $M|D$, choose $z$ maximizing $z(D)$ subject to $z\le\alpha x|_D$. Let $A$ be the union of all sets whose rank constraints are tight at $z$. By submodularity, $A$ is tight, so $z|_A$ witnesses that $A$ can be made tight. Moreover, $z_e=\alpha x_e$ for every $e\in D\setminus A$, since otherwise we could increase $z_e$.

		To see that $A$ contains every set $B\subseteq D$ that can be made tight, choose $w\in\PB{M|B}$ with $w\le\alpha x|_B$. By submodularity and the rank constraint for $A\cap B$ at $w$, we have
		\[
			r_M(A\cup B)-r_M(A)
			\le r_M(B)-r_M(A\cap B)
			\le w(B\setminus A)
			\le \alpha x(B\setminus A)=z(B\setminus A).
		\]
		Since $A$ is tight at $z$, this gives $r_M(A\cup B)\le z(A\cup B)$. Thus $A\cup B$ is also tight at $z$, so $B\subseteq A$ by the definition of $A$. We can therefore take $A(D)=A$. The asserted bound follows from $r_M(A)+\alpha x(D\setminus A)=z(D)\le r_M(D)$.

		Finally, greedy optimization gives polynomial-time separation for $\PI{M|D}$, so we can compute $z$ by linear programming. To recover $A$, test for each $e\in D$ whether $z_e$ can be increased in $\PI{M|D}$ with all other coordinates fixed, ignoring the upper bound $\alpha x_e$ in this test. Exactly the elements in a tight set fail the test, so these elements form $A$.
	\end{proof}
	
	\begin{proof}[Proof of \cref{claim:bounded_refinement_point}]
		Since $N$ is a direct sum, it suffices to find a point in each $\PB{N_i}$ bounded by $\alpha x|_{E_i}$. Fix $i$ and write $A=A(D_i)$. Since $A$ can be made tight, there is a point $w\in\PB{M|A}$ with $w\le\alpha x|_A$. Write $w$ as a convex combination of incidence vectors of bases of $M|A$.

		Consider any basis $B$ in this combination. Since $B$ spans $A$, the set $C_{i-1}\cup B$ spans $C_i=C_{i-1}\cup A$. Thus, after contracting $C_{i-1}$, the set $B\setminus C_{i-1}$ spans $N_i$ and contains a basis $B_i$ of $N_i$. Replacing each $B$ by such a $B_i$ with the same coefficient gives a point $y_i\in\PB{N_i}$. Since $B_i\subseteq B$, we have $y_i\le w|_{E_i}\le\alpha x|_{E_i}$. Combining these points proves the claim.
	\end{proof}
	This finishes the proof of \cref{lem:partial_alignment_matroid}.
\end{proof}

\subsubsection{Iterative relaxation for weakly aligned points}

It remains to prove \cref{lem:rounding_thin_trees}. For $S\subseteq V$, let $\bar E[S]$ denote the edges of the original graph with both endpoints in $S$.
For each $L\in\mathcal L$, we count edges in $\delta(L)$ once and edges in $\bar E[L]\setminus D_L$ twice. This lets us compare the load of a parent with the loads of its children. The two steps of the algorithm are illustrated in \cref{fig:parent-relaxation}. If an edge has value at least $\tfrac12$ in the current extreme point, we round it up as in \cref{fig:parent-relaxation}~(a). Otherwise, we find a tight constraint for a set $L$ that counts every edge with at least the same multiplicity as its children combined, as in \cref{fig:parent-relaxation}~(b). The extra edge multiplicities counted by $L$, shown in orange, have average fractional value at least $\tfrac14$. Even if the basis takes them all, they contribute at most four times their fractional load. We can therefore drop the constraint for $L$ and use the bounds for its children to control the rest of its load.

\begin{proof}[Proof of \cref{lem:rounding_thin_trees}]
	For $L\in\mathcal L$ and $e\in E$, define
	\[
		a_L(e)=
		\begin{cases}
			1 & \text{if } e\in\delta(L),\\
			2 & \text{if } e\in \bar E[L]\setminus D_L,\\
			0 & \text{otherwise.}
		\end{cases}
	\]
	For $x\in\mathbb R^E$ and $B\subseteq E$, write
	\[
		a_L(x)=\sum_{e\in E}a_L(e)x_e,\quad \text{and} \quad a_L(B)=\sum_{e\in B}a_L(e).
	\]
	Set $b_L=\lceil 2a_L(y)\rceil/2$.
	Let $x^*$ be a vertex of the polytope 
	\begin{equation}\label{eq:constrained_polytope}
		\{x \in \PB{M} : a_L(x) \le b_L\quad\forall L\in\mathcal L\}.
	\end{equation}
	We show by induction on $|E|+|\mathcal{L}|$ that we can round $x^*$ to a basis $B$ of $M$ with $a_L(B) \le 4b_L$ for each $L \in \mathcal{L}$, yielding the desired bound since $b_L \le a_L(y)+\frac12$.
	
	On deleting or contracting an edge, we retain the coefficients $a_L(f)$ of all remaining edges. 
	In the induction, we require the budgets to be half-integral. 
	In the sum $M=\bigoplus_{i=1}^k M_i$, we only keep matroids with nonempty ground set.
	For each such summand $M_i$, let $\bar E_i$ denote its original ground set and $E_i$ its current ground set.
	
	If $E=\emptyset$, choose $B = \emptyset$.
	Note that if $x^*_e = 0$ for some $e \in E$, we can delete $e$ and directly apply induction.
	
	Otherwise, we round $x^*$ using iterative rounding and relaxation.
	More precisely, we show that either there is an edge $e \in E$ with $x^*_e\ge\frac12$, in which case we round up $x^*_e$, or we can relax a constraint in $\mathcal{L}$.
	
	First, assume $x^*_e\ge\frac12$ for some $e \in E$.
	Let $b'$ result from $b$ by reducing $b_L$ by one half for each $L \in \mathcal{L}$ with $a_L(e)>0$.
	Delete $e$ from $E$, and let $M' = M / \{e\}$ be the minor of $M$ on ground set $E \setminus \{e\}$. 
	Note that \eqref{eq:constrained_polytope} remains nonempty as there is some $x' \in \PB{M'}$ with $x'_{f} \le x^*_{f}$ for each $f \in E \setminus \{e\}$.
	By induction, we can compute a basis $B'$ of $M'$ with $a_L(B') \le 4b'_L$ for each $L \in \mathcal{L}$.
	Thus, $B=B' \cup \{e\}$ is a basis of $M$ with
	\[
	a_L(B) \le 4b'_L + a_L(e) \le 4b_L
	\]
	for each $L \in \mathcal{L}$.
	
	If $x^*_e < \frac12$ for each $e \in E$, we can identify a constraint in $\mathcal{L}$ to relax as follows.
	Let $\mathcal{U} \subseteq 2^E$ and $\mathcal{L}' \subseteq \mathcal{L}$ with $|\mathcal{U}|$ maximal such that the rank inequalities on $\mathcal{U}$ and the constraints $a_L(x) \le b_L$ for $L \in \mathcal{L}'$ are tight and linearly independent and uniquely determine $x^*$.
	For $L\in\mathcal L'$ with children $L_1,\dots,L_h$ in $\mathcal L'$, let
	\begin{equation}\label{eq:multiplicity}
		n_L(e)=a_L(e)-\sum_{i=1}^h a_{L_i}(e)
	\end{equation}
	be the difference between the parent’s coefficient and the sum of its children’s coefficients.
	Define $n_L(x)$ and $n_L(B)$ by summing these coefficients as for $a_L$.
	\begin{claim}\label{claim:relax}
		There is a set $L\in\mathcal L'$ such that $n_L(e)\ge0$ for every $e\in E$, $n_L(E)>0$, and
		\[
			\frac{n_L(x^*)}{n_L(E)}\ge\frac14.
		\]
	\end{claim}
	Before proving \cref{claim:relax}, let us conclude the proof of \cref{lem:rounding_thin_trees}.
	Let $L \in \mathcal{L}'$ be the set given by \cref{claim:relax}.
	We relax $L$ and apply induction. 
	This gives a basis $B$ of $M$ with $a_{L'}(B) \le 4b_{L'}$ for each $L' \in \mathcal{L} \setminus \{L\}$.
	It remains to show $a_L(B) \le 4b_{L}$.
	Note that
	\begin{align*}
	a_L(B) &= \sum_{i=1}^h a_{L_i}(B)+n_L(B) \\
	&\le \sum_{i=1}^h 4 b_{L_i} + 4n_L(x^*) \\
	&= \sum_{i=1}^h 4 a_{L_i}(x^*) + 4n_L(x^*) \\
	&= 4a_L(x^*) \\
	&= 4 b_L.
	\end{align*}
	The inequality uses induction for the children of $L$ and $n_L(B)\le n_L(E)\le4n_L(x^*)$ by \cref{claim:relax}. The remaining equalities follow from tightness of sets in $\mathcal{L}'$ and \eqref{eq:multiplicity}.
	
	To find such a constraint to relax in polynomial time, we test for each $L\in\mathcal L$ whether
	$a_L(z)\le 4b_L$ is implied by $z\in[0,1]^E$ and
	$a_{L'}(z)\le 4b_{L'}$ for all $L' \in\mathcal L\setminus\{L\}$.
	Each test is a linear program that maximizes $a_L(z)$, and the preceding argument shows that
	the set $L$ given by \cref{claim:relax} passes this test,
	since $n_L(z)\le n_L(E)$ for every $z\in[0,1]^E$.
	We may relax any constraint that passes.
	
	Thus, it remains to prove \cref{claim:relax}.
	\begin{proof}[Proof of \cref{claim:relax}]
		Let $m=|\mathcal U|$ and $\ell=|\mathcal L'|$.
		Call $L\in\mathcal L'$ exceptional if $n_L(e)<0$ for some $e\in E$, and let $b$ be the number of exceptional sets. We first show that $b<m$.

		For $e\in E_i$ with $i \in [k]$, if $\bar E_i\not\subseteq \bar E[L]$, then $a_L(e)$ counts the endpoints of $e$ in $L$, and the same holds for its children. In this case $n_L(e)$ counts the endpoints in $L\setminus\bigcup_j L_j$. Otherwise $a_L(e)=0$, so $n_L(e)\le0$. Consequently, if $n_L(e)<0$, then $\bar E_i\subseteq \bar E[L]$ and some child of $L$ in $\mathcal{L}'$ counts $e$. 
		This implies that $L$ is the smallest member of $\mathcal L'$ with $\bar E_i\subseteq \bar E[L]$. Distinct exceptional sets therefore have distinct witnessing summands $M_i$.
		As all summands have nonempty ground set, at least one summand has $\bar E_i\not\subseteq \bar E[R]$ for every maximal $R\in\mathcal L'$. Otherwise every maximal $R$ would have $a_R(e)=0$ for every $e \in E$, contrary to linear independence.
		This implies $b < k$.
		
		Note that $k \le m$ by maximality of $m$. Indeed, the nonempty ground sets $E_1,\dots,E_k$ of the matroids give linearly independent tight rank constraints.
		This proves $b<k\le m$.
		
		Now consider a nonexceptional set $L \in \mathcal{L}'$ with children
			$L_1,\dots,L_h$ in $\mathcal{L}'$. By \eqref{eq:multiplicity} and tightness,
		\[
		n_L(x^*)=b_L-\sum_{i=1}^{h} b_{L_i}
			\in\tfrac12\mathbb Z.
		\]
		Moreover, $n_L(E)>0$ by linear independence.
			Since $0<x_e^*<1/2$ for each $e\in E$,
			$n_L(E)-2n_L(x^*)$ is a positive integer, and $n_L(E) \ge 2$. 
			If $n_L(E) - 2n_L(x^*) = 1$, then 
			\[
			\frac{n_L(x^*)}{n_L(E)}
			=\frac{n_L(E)-1}{2n_L(E)}\ge\frac14,
			\]
			as desired.
		So suppose $n_L(E)-2n_L(x^*) \ge 2$ for each nonexceptional set $L \in \mathcal{L}'$. 
			For a nonexceptional $L$, the description of $n_L(e)$ above shows that it is at most the number of endpoints of $e$ in $L\setminus\bigcup_j L_j$. These regions are pairwise disjoint over $L\in\mathcal L'$, so the sum of $n_L(e)$ over the nonexceptional sets is at most two. This gives
		\begin{equation}\label{eq:deficit_lb}
			2(\ell-b)
				\le \sum_{\substack{L\in\mathcal L'\\L\text{ nonexceptional}}}
				\Bigl(n_L(E)-2n_L(x^*)\Bigr)
				\le 2\bigl(|\supp(x^*)|-2x^*(E)\bigr),
		\end{equation}
		where in the first inequality, we used $n_L(E)-2n_L(x^*) \ge 2$ for each nonexceptional set $L \in \mathcal{L}'$. 
		In the second inequality, we used the above counting argument and that $1-2x_e^*>0$ for each $e \in E$.
		
			By a standard uncrossing argument, we can assume the sets in $\mathcal{U}$ to form a chain $U_1 \subsetneq \ldots \subsetneq U_m$. Using $x^*(U_{i}) \ge x^*(U_{i-1})+1$ for each $i \in [m]$ (with $U_0 := \emptyset$), we get
			\begin{equation*}
				|\supp(x^*)|-2x^*(E)
				=m+\ell-2x^*(E)\le \ell-m.
			\end{equation*}
		Together with \eqref{eq:deficit_lb}, this gives $b\ge m$, a contradiction.
	\end{proof}
	This finishes the proof of \cref{lem:rounding_thin_trees}.
\end{proof}

\ifbool{ipcosubm}{}{\section*{AI acknowledgement}
We used ChatGPT-6 for assisting with technical statements and proofs, and also to improve the write-up.
\printbibliography
\appendix
}

\section{Polyhedral interpretation of matroid refinements}
\label{sec:refinement}

For completeness, we provide a short proof verifying that each nonempty face of a matroid base polytope is itself a matroid base polytope.

\begin{lemma}\label{lem:refinement}
	Let $M$ be a matroid on ground set $N$. 
	Let $F$ be a nonempty face of $\PB{M}$.
	Then, there is a matroid $M'$ on ground set $N$ with $\PB{M'} = F$. 
	Moreover, if $M$ is given by a polynomial-time independence oracle, and $F$ is specified as $F_{\PB{M}}(x)$ for a given point $x\in\PB{M}$, then we can efficiently construct a polynomial-time independence oracle for $M'$.
\end{lemma}

The latter statement is used in the proof of \cref{lem:matroid_equivalence2}.

\begin{proof}
	We start with showing existence of $M'$.
	Choose $x$ in the relative interior of $F$, and let
	$\mathcal{T}:=\{S\subseteq N:x(S)=r_M(S)\}$.
	Submodularity implies that $\mathcal{T}$ is closed under unions
	and intersections.
	Choose a maximal chain
	$\emptyset=C_0\subsetneq\ldots\subsetneq C_k=N$
	in $\mathcal{T}$, and define
	\[
	M':=\bigoplus_{i=1}^k M_i
	\]
	with $M_i := (M|C_i)/C_{i-1}$ obtained by restricting $M$ to $C_i$ and then contracting $C_{i-1}$.
	The bases of $M'$ are precisely the bases $B$ of $M$ satisfying
	$|B\cap C_i|=r_M(C_i)$ for every $i \in [k]$.
	
	For any $T\in\mathcal{T}$, the set
	$C_{i-1}\cup(T\cap C_i)$ belongs to $\mathcal{T}$.
	By maximality of the chain, it equals either $C_{i-1}$ or $C_i$.
	Thus every $T\in\mathcal{T}$ is a union of blocks
	$C_i\setminus C_{i-1}$, so for each basis $B$ of $M'$, we have
	\begin{equation}\label{eq:bases_of_refinement}
	\begin{aligned}
	|B\cap T|
	&= \sum_{i \in [k] : C_i \setminus C_{i-1} \subseteq T} |B \cap (C_i \setminus C_{i-1})| \\
	&= \sum_{i \in [k] : C_i \setminus C_{i-1} \subseteq T} (r_M(C_i) - r_M(C_{i-1})) \\
	&= r_M(T).
	\end{aligned}
	\end{equation}
	For the last equality, note that $r_M(C_i)-r_M(C_{i-1})=x(C_i\setminus C_{i-1})$
	since $C_i,C_{i-1}\in\mathcal T$. The blocks contained in $T$ partition $T$,
	so these differences sum to $x(T)=r_M(T)$.
	\eqref{eq:bases_of_refinement} implies $e\notin B$ whenever $x_e=0$, since otherwise 
	\[
	|B| > |B \cap (N \setminus \{e\})| = r_M(N \setminus \{e\}) = r_M(N) = |B|,
	\]
	where we used \eqref{eq:bases_of_refinement} with $T = N \setminus \{e\}$ in the first equality. 
	Indeed, $N \setminus \{e\} \in \mathcal{T}$ as $x_e = 0$.
	
	Hence, for each basis $B$ of $M'$, every defining inequality of $\PB{M}$ that is tight at $x$ is also tight at $\chi^B$, so $\chi^B\in F$.
	Conversely, if $B$ is a basis of $M$ with $\chi^B\in F$, then $|B\cap C_i|=r_M(C_i)$ for every $i\in[k]$, since each of these rank inequalities is tight at $x$ and hence throughout $F$. Thus $B$ is a basis of $M'$.
	The bases of $M'$ are therefore exactly the bases in $F$,
	and taking convex hulls yields $\PB{M'}=F$.

It remains to construct an independence oracle for $M'$.
We first compute a maximal chain in $\mathcal T$ for the given point $x$.
Start with $\emptyset$ and $N$. For two consecutive sets $A\subsetneq B$ in the current chain and each ordered pair of distinct elements $e,f\in B\setminus A$, minimize
$r_M(S)-x(S)$ over $A\cup\{e\}\subseteq S\subseteq B\setminus\{f\}$.
This is a submodular function minimization problem on the free elements and can be solved in polynomial time \cite{Schrijver03}, since we can evaluate $r_M$ by the greedy algorithm using the independence oracle for $M$.
If the minimum is zero, insert the minimizing set between $A$ and $B$ and repeat.
Indeed, any $T\in\mathcal T$ with $A\subsetneq T\subsetneq B$ gives a minimum of zero for $e\in T\setminus A$ and $f\in B\setminus T$, since $r_M(S)-x(S)\ge0$ for all $S\subseteq N$.
Thus the chain is maximal when no further insertion is possible.
Since the chain has at most $|N|+1$ members and we test at most
$|N|^2$ ordered pairs per iteration, the construction takes polynomial time.

Given $J\subseteq N$, let $J_i=J\cap(C_i\setminus C_{i-1})$. Since $M'$ is the direct sum of the minors $(M|C_i)/C_{i-1}$, $J$ is independent in $M'$ if and only if
\[
 r_M(C_{i-1}\cup J_i)-r_M(C_{i-1})=|J_i| \qquad \forall \, i\in[k].
\]
These rank queries give the desired polynomial-time independence oracle.
\end{proof}
\section{From exact cost to exact marginals}
\label{app:marginals}

From the exact-cost guarantee for laminar-thin trees given by \Cref{thm:intro-laminar}, one readily obtains through standard techniques the following result.

\begin{corollary}\label{cor:marginals}
Given $G$, $\cL$, and $x$ as in \cref{thm:intro-laminar}, there is a polynomial-time algorithm that computes a probability distribution on spanning trees such that
\[
  \Pr[e\in T]=x_e\qquad\forall\; e\in E.
\]
Every supported tree is $26$-thin for $\cL$ with respect to $x$ and lies in $F_Q(x)$, where $Q$ denotes the spanning tree polytope on $G$.
\end{corollary}

\begin{proof}[Proof of \cref{cor:marginals}]
Let $\mathcal T\subseteq2^E$ be the finite family of all spanning-tree edge sets that are 26-thin for $\cL$ with respect to $x$.
We construct the distribution by collecting polynomially many trees whose incidence vectors have $x$ in their convex hull.
Consider the feasibility system in the variable $c\in\mathbb{R}^E$
\[
  c^{\mathsf T}(\chi^T-x)\ge1\qquad\forall\; T\in\mathcal T.
\]
Note that for a feasible cost vector $c$ we have $c(T) \ge c^\top x + 1$ for each $26$-thin tree. 
Thus, this system is infeasible by \cref{thm:intro-laminar}, which guarantees for every cost vector $c$ a tree $T\in\mathcal T$ with $c^{\mathsf T}(\chi^T-x)\le0$.

We solve the separation problem for this system as follows. Given a
query point $c'$, apply \cref{thm:intro-laminar} to obtain a tree
$T\in\mathcal T$ with $c'^\top(\chi^T-x) \le 0$. If $\chi^T=x$, we return
the distribution assigning probability one to $T$. Otherwise, we return
the inequality $c^\top(\chi^T-x)\ge 1$: every feasible $c$ must satisfy
it, while the query point $c'$ violates it, and its normal $\chi^T-x$ is nonzero.

Run the ellipsoid feasibility algorithm~\cite[Theorem 6.4.1]{GLS93} with
this oracle and record the trees $\mathcal T_0$ it returns. Unless we have already found
the desired distribution, the algorithm certifies infeasibility after
polynomially many queries. 

The inequalities of $\mathcal T_0$ alone
form an infeasible subsystem, since the same oracle answers give the same
ellipsoid run on that subsystem.
Consequently, $x$ lies in the convex hull of the recorded incidence vectors, since otherwise strict separation, rescaled to margin one, would give a feasible $c$ for the subsystem.
We can therefore compute the probability vector $\lambda\in\mathbb{R}_{\ge0}^{\mathcal T_0}$ by solving the explicit linear program
\[
  \sum_{T \in \mathcal T_0}\lambda_T\chi^T=x,\qquad
  \sum_{T \in \mathcal T_0}\lambda_T=1,\qquad \lambda_T\ge0.
\]
There are polynomially many variables and constraints, so this computation takes polynomial time.
Every supporting inequality of the spanning tree polytope that is tight at $x$ must also be tight at every tree with positive probability: all its slacks are nonnegative and their average is zero.
Such trees therefore lie in $F_Q(x)$, as required.
\end{proof}

\end{document}